\documentclass[namedreferences]{solarphysics}

\usepackage[hyperref,optionalrh]{spr-sola-addons} 
\usepackage{graphicx}        
\usepackage{color}           
\usepackage{breakurl}        
\usepackage{mathptmx}        

\hypersetup{
    final=true,
    pageanchor=true,
    colorlinks=true,
    breaklinks=true,
    linkcolor=blue,
    citecolor=blue,
    urlcolor=blue,
    pdfpagemode=UseNone,
    pdftitle={},
    pdfauthor={},
    pdfsubject={},
    pdfkeywords={}}

\def\arcsec{\hbox{$^{\prime\prime}$}}

\chardef\us=`\_

\definecolor{myBlue}{rgb}{.2,.7,.9}
\definecolor{myDarkRed}{rgb}{0.698, 0.094, 0.133}

\newcommand{\kms}{km\,s$^{-1}$}

\begin{document}

\begin{article}
\begin{opening}

\title{Magnetic Flux Emergence in a Coronal Hole}

\author[addressref={aff1, aff2}, corref,email={jpalacios@leibniz-kis.de, j.palacios@rheagroup.com}]{\inits{J.}\fnm{Judith}~\lnm{Palacios}{\orcid{0000-0002-1518-512X}}}
\author[addressref={aff3,aff4},corref, email={dominik.utz@uni-graz.at}]{\inits{D.}\fnm{Dominik}~\lnm{Utz}\orcid{0000-0002-0061-5916}}
\author[addressref={aff3}]{\inits{S.}\fnm{Stefan}~\lnm{Hofmeister}}
\author[addressref={aff3}]{\inits{K.}\fnm{Kilian}~\lnm{Krikova}\orcid{0000-0002-0922-7864}}
\author[addressref={aff5}]{\inits{P.}\fnm{Peter}~\lnm{G\"om\"ory}\orcid{0000-0002-0473-4103}}
\author[addressref={aff6}]{\inits{C.}\fnm{Christoph}~\lnm{Kuckein}\orcid{0000-0002-3242-1497}}
\author[addressref={aff6}]{\inits{C.}\fnm{Carsten}~\lnm{Denker}\orcid{0000-0002-7729-6415}}
\author[addressref={aff6}]{\inits{M.}\fnm{Meetu}~\lnm{Verma}\orcid{0000-0003-1054-766X}}
\author[addressref={aff5}]{\inits{S. J.}\fnm{Sergio Javier}~\lnm{Gonz\'alez Manrique}\orcid{0000-0002-6546-5955}}
\author[addressref={aff3}]{\inits{J.~I.}\fnm{Jose Iv\'an}~\lnm{Campos Rozo}\orcid{0000-0001-8883-6790}}
\author[addressref={aff5}]{\inits{J.}\fnm{J\'{u}lius}~\lnm{Koza}\orcid{0000-0002-7444-7046}}
\author[addressref={aff3}]{\inits{M.}\fnm{Manuela}~\lnm{Temmer}}
\author[addressref={aff3}]{\inits{A.}\fnm{Astrid}~\lnm{Veronig}}
\author[addressref={aff6,aff7}]{\inits{A.}\fnm{Andrea}~\lnm{Diercke}\orcid{0000-0002-9858-0490}}
\author[addressref={aff6}]{\inits{I.}\fnm{Ioannis}~\lnm{Kontogiannis}\orcid{0000-0002-3694-4527}}
\author[addressref={aff8}]{\inits{C.}\fnm{Consuelo}~\lnm{Cid}\orcid{0000-0002-2863-3745}}

\address[id=aff1]{Leibniz-Institut f\"ur Sonnenphysik (KIS), Sch\"oneckstra{\ss}e 6, 79104 Freiburg, Germany}
\address[]{$^{2}$ RHEA Group, Robert-Bosch-Str. 7, 64293 Darmstadt, Germany}
\address[id=aff3]{IGAM, Institute of Physics, University of Graz, Austria}
\address[id=aff4]{IAA-CSIC, Instituto de Astrof{\'{\i}}sica de Andaluc{\'{\i}}a, Granada, Spain}
\address[id=aff5]{Astronomical Institute, Slovak Academy of Sciences, Tatransk\'a Lomnica, Slovakia}
\address[id=aff6]{Leibniz-Institut f\"ur Astrophysik  (AIP), An der Sternwarte 16, 14482 Potsdam, Germany}
\address[id=aff7]{Universit\"at Potsdam, Institut f\"ur Physik und Astronomie, Karl-Liebknecht-Stra{\ss}e 24-25, 14476 Potsdam, Germany}
\address[id=aff8]{Space Weather Group, Dpto de F\'isica y Matem\'aticas, Universidad de Alcal\'a, Madrid, Spain}

\runningauthor{J. Palacios et al.}
\runningtitle{Magnetic Flux Emergence in a Coronal Hole}

\begin{abstract}
A joint campaign of various space-borne and ground-based observatories, comprising the Japanese Hinode mission (HOP~338, 20\,--\,30~September 2017), the GREGOR solar telescope, and the \textit{Vacuum Tower Telescope} (VTT), investigated numerous targets such as pores, sunspots, and coronal holes. In this study, we focus on the coronal hole region target. On 24~September 2017, a very extended non-polar coronal hole developed patches of flux emergence, which contributed to the decrease of the overall area of the coronal hole. These flux emergence patches erode the coronal hole and transform the area into a more quiet-Sun-like area, whereby bipolar magnetic structures play an important role. Conversely, flux cancellation leads to the reduction of opposite-polarity magnetic fields and to an increase in the area of the coronal hole.

Other global coronal hole characteristics, including the evolution of the associated magnetic flux and the aforementioned area evolution in the EUV, are studied using data of the \textit{Helioseismic and Magnetic Imager} (HMI) and \textit{Atmospheric Imaging Assembly} (AIA) onboard the \textit{Solar Dynamics Observatory} (SDO). The interplanetary medium parameters of the solar wind display parameters compatible with the presence of the coronal hole. Furthermore, a particular transient is found in those parameters.

\end{abstract}

\end{opening}

\section{Introduction}
     
Coronal holes are relatively large domains of the outer solar atmosphere with significantly weakened emissivity in the EUV spectral range due to a somewhat reduced low density and predominantly open magnetic field structure with a slight dominance of one polarity. A comprehensive review on the subject is given in, \textit{e.g.}, \citet{Wang2009}. Seminal observations by \citet{Altschuler1972} targeted the coronal areas. While some coronal holes are formed as extensions of polar coronal holes \citep[\textit{e.g.}][]{Harvey2002}, coronal holes can evolve by themselves in non-polar regions at low heliographic latitudes. They can also form as a consequence of long-term decay of active regions \citep{Karachik2010} or after eruptive events, which can trigger an opening of the field lines \citep{Heinemann2018a}. The latter study also addresses differential rotation of coronal holes, which is largest at the equator likely due to conservation of angular momentum. Coronal holes are the source of the fast solar wind, which is related to high-speed streams in the interplanetary medium \citep{Krieger1973}. When they meet with a slow stream interaction regions are formed, which are observed and identified in {\it in-situ} measurements as corotating interacting regions.

Coronal holes are faint structures. Therefore, they were subjects of more detailed identification methods only recently \citep{Krista2009} accompanied by more systematic analysis and cataloging effort such as the \textsf{Spatial Possibilistic Clustering Algorithm} \citep[SPoCA,][]{Verbeek2014} and the \textsf{Collection of Analysis Tools for Coronal Holes} \citep[CATCH,][]{Heinemann2019}. Coronal holes can be identified and extracted either by supervised \citep{Reiss2015} or unsupervised methods \citep{Arish2016}.
  
Important characteristics of low-latitude coronal holes are summarized in \citet{Hofmeister2017} such as predominant polarity and correlations between area, magnetic flux, and latitudes. These detailed analyses show that coronal holes are composed of flux tubes of one dominant polarity rooted mostly in the network. The evolution of a coronal hole from birth to decay is followed in \citet{Heinemann2018a,Heinemann2018b} together with parameters of interplanetary medium. \citet{Hofmeister2019} show statistics of coronal holes and estimate their predominant magnetic flux. 

A relatively unexplored domain is the relation between large-scale structure of magnetic fields associated with coronal holes and the smallest detectable flux concentrations referred to as magnetic elements or magnetic bright points. Their properties, extensively studied in, \textit{e.g.}, \citet{Utz2014} and \citet{Bodnarova2014}, may provide important clues for understanding the global behavior of coronal holes.    
 
In this work, we study a coronal hole, which exhibits several emergence processes of small- and medium-scale flux concentrations in the course of several days. Flux emergence is a broad topic, which is not commonly treated  in the context of associated interplanetary events. A first attempt was made in \citet{Navarro2019} showing simulation results of flux emergence inside a coronal hole.

Some particular cases of flux emergence within coronal holes are the so called ``anemones'' \citep{Asai2008}, which are formed by active regions that grow inside a coronal hole, showing a potential appearance. Some cases of anemone growth and related geomagnetic disturbances were reported by \citet{Asai2009} and \citet{Cid2014}. However, this relation was due to the associated coronal mass ejections (CMEs) originating in newly emerging active regions -- sometimes of puzzling complexity -- and not due to flux emergence itself in the coronal hole. These regions are similar to those flux emergence cases registered as ephemeral regions \citep{Harvey1973}. The regions in this study are named emerging flux regions (EFRs), as in \citet{Kontogiannis2019}, similar to those found on smaller scales in the quiet Sun.
 
Magnetic flux cancellation was also investigated with respect to coronal holes, and more specifically to their boundaries in relation to jet production \citep{MorenoInsertis2008}, even though these features are found everywhere in the Sun \citep{Shibata2007}.
 
The coronal hole shown in this study exhibits different flux emergence occurrences over the course of several days. These flux emergence events lead to local cancellation, brightening, formation of X-ray bright points (XBPs), and an overall shrinking of the coronal hole area due to the flux emergence. This last property is sought in interplanetary data, related with the solar wind speed, densities, and charge states composition.  
 
This study is organized as follows: the datasets are described in Section~\ref{data_sec}, the coronal hole morphology and its characteristics are introduced in Section~\ref{CH_sec}. The flux emergence events are presented and analysed in the Sections~\ref{ROIs_sec} and~\ref{emergcancel_sec}. The interplanetary medium parameters associated with the coronal hole are analysed in Section~\ref{IP_sec}. Finally, Sections~\ref{disc_sec} and \ref{concl_sec} contain a discussion of the results and concluding remarks, respectively.

\section{Observations}\label{data_sec}

The data in this work were acquired with the \textit{Solar Dynamics Observatory} \citep[SDO,][]{Pesnell2012}. Time-series of EUV full-disk images with 30-minute cadence were obtained with the \textit{Atmospheric Imaging Assembly} \citep[AIA,][]{Lemen2012}, and full-disk data of the \textit{Helioseismic and Magnetic Imager} \citep[HMI,][]{Scherrer2012} were used to track the temporal evolution of area and flux, respectively. Datasets were reduced with standard procedures. The EUV images were recorded in the AIA passbands 193\,\AA\ and 211\,\AA. In coronal holes the main contributors to the former passband are the emission lines \mbox{Fe\,\textsc{xii}}~195.12\,\AA, \mbox{Fe\,\textsc{xii}}~193.51\,\AA, and   \mbox{Fe\,\textsc{xxi}}~188.23\,\AA, whereas the dominant contribution to the latter comes from \mbox{Fe\,\textsc{xxi}}~209.78\,\AA\ \citep{ODwyer2010}. 
The Hinode Observing Plan (HOP) 338 was executed from 20~September to 30~Sep\-tem\-ber 2017 delivering detailed data from the \textit{X-Ray Telescope} \citep[XRT,][]{Golub2007} on board the Hinode spacecraft \citep{Kosugi2007}. The XRT 1-minute cadence Al\_mesh data and corresponding HMI 45-second cadence data are of particular interest because they feature emerging XBPs on 26~September 2017. The XRT datasets were reduced by standard procedures.
 
Weather conditions at the GREGOR solar telescope \citep{Schmidt2012} and the \textit{Vacuum Tower Telescope} \citep[VTT,][]{vonderLuehe1998}, both located at the Observatorio del Teide, Tenerife, Spain, prevented capturing high-resolution data. The only datasets from that period were acquired with the \textit{Chromospheric Telescope} \citep[ChroTel,][]{Kenti2008, Bethge2011} operated at VTT. We use ChroTel filtergrams because coronal holes are well observed in the \mbox{He\,\textsc{i}}~10830\,\AA\ triplet due to reduced absorption in the chromosphere \citep{Harvey2002}. The ChroTel images were corrected for intensity variations introduced by non-uniform filter transmission as described in \citet{Shen2018} and \citet{Diercke2019}.
 
Taking advantage of this property of the \mbox{He\,\textsc{i}}~10830\,\AA\ triplet, near-infrared Stokes-$V/I$ full-disk magnetograms from the \textit{Solar Flare Telescope} (SFT) at Mitaka, Japan\footnote{\href{http://solarwww.mtk.nao.ac.jp/en/solarobs.html}{solarwww.mtk.nao.ac.jp/en/solarobs.html}} were included in this study. In addition, Stokes-$V/I$ full-disk magnetograms are available for the \mbox{Si\,\textsc{i}}~10827\,\AA\ and \mbox{Fe\,\textsc{i}}~15658\,\AA\ photospheric spectral lines \citep{Sakurai2018}.
 
\section{Coronal Hole Morphology and Characteristics}\label{CH_sec}

In HOP~338, one of the targets was a large non-polar coronal hole, covering the latitude range from $-20^{\circ}$\,S to $60^{\circ}$\,N, where it joins a polar coronal hole. This equatorial coronal hole is long-lived, \textit{i.e.}, it appeared about three solar rotations before the start of HOP~338 in SDO observations on 9~June 2017 and in Stereo~A \citep{Kaiser2008} observations on 25~June 2017. Thus, the coronal hole was already well established for the observation time. The coronal hole lived for at least another three rotations, before becoming more roundish first and completely deformed thereafter. 
 
\begin{figure}
\includegraphics[width=0.30\textheight,clip=]{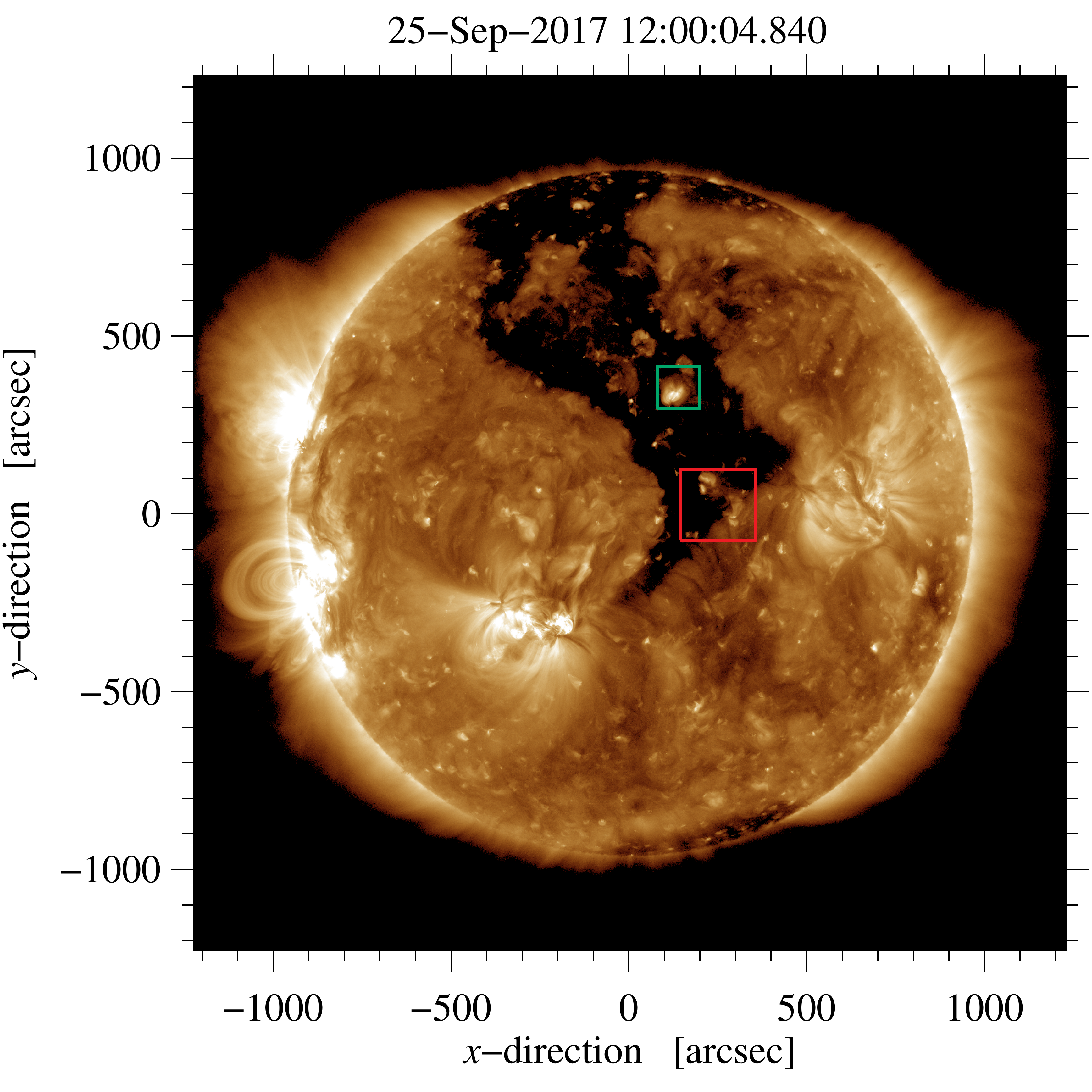}
\includegraphics[width=0.3\textheight,clip=]{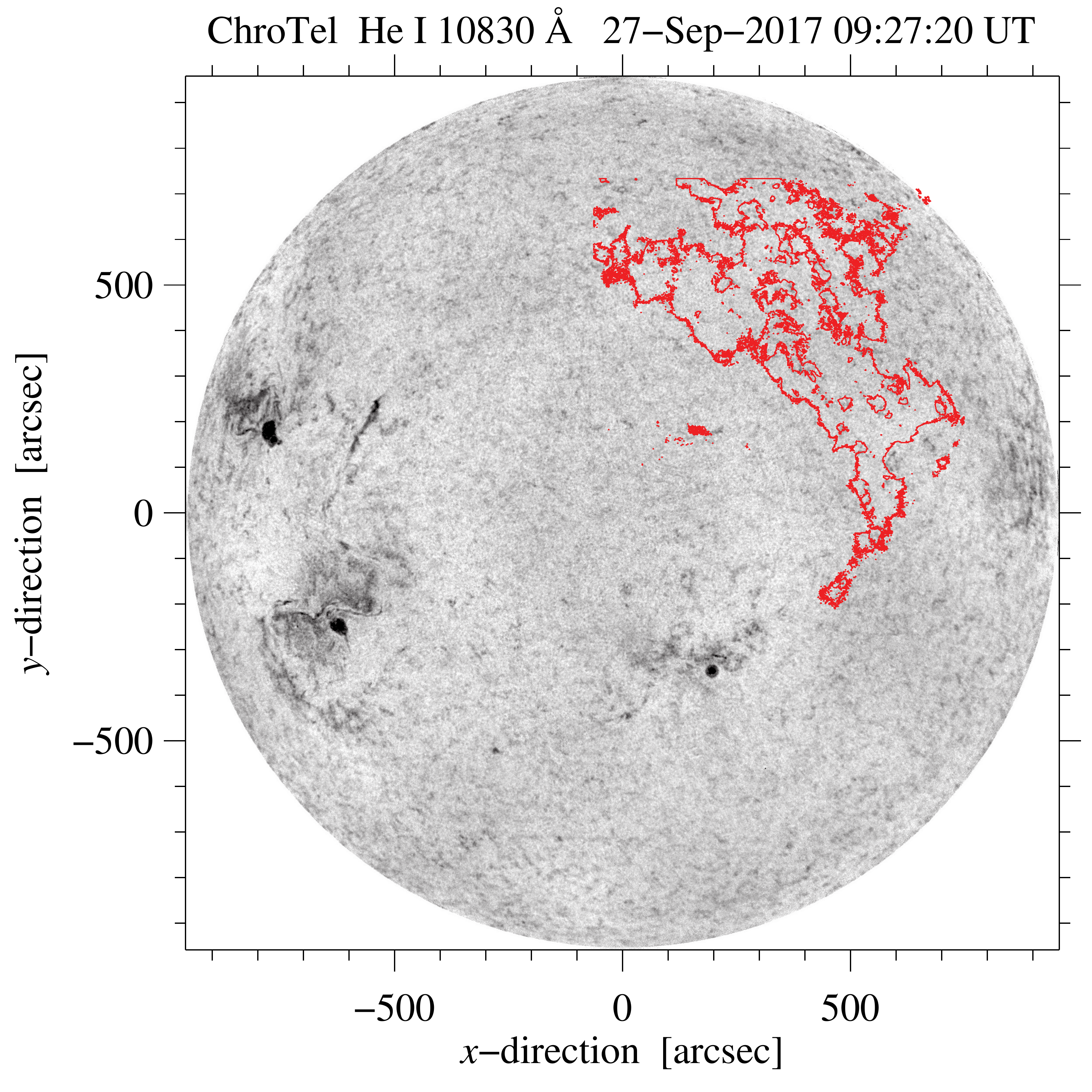}
\caption{Full-disk AIA EUV 193\,\AA\ image on 25~September 2017 (\textit{left}) and ChroTel He\,\textsc{i} 10830\,\AA\ line-core image on 27~September 2017 (\textit{right}), providing context information for the coronal hole studies. The  green and red boxes refer to ROI\,1 and ROI\,2, respectively. The red contours outline the coronal hole area.}
\label{context_aia_chrotel}
\end{figure}

We show a context EUV 193\,\AA\ image acquired on 25~September 2017 in Figure~\ref{context_aia_chrotel}, which depicts and outlines the coronal hole. A full-disk enhanced 193\,\AA\ movie which encompasses several days is included as Supplementary Material and explained in the Appendix. A green box encloses a region-of-interest (ROI) with a flux emergence event, which is referred to as ROI\,1. Similarly, a larger red box encloses ROI\,2 with a number of flux emergence events. The flux emergence events are located at the center of the coronal hole on 25~September 2017 in ROI\,1 and at its periphery on 26~September 2017 in ROI\,2. The events of in both ROIs are described and analysed in Section~\ref{ROIs_sec}.

\begin{figure}
\includegraphics[width=\textwidth]{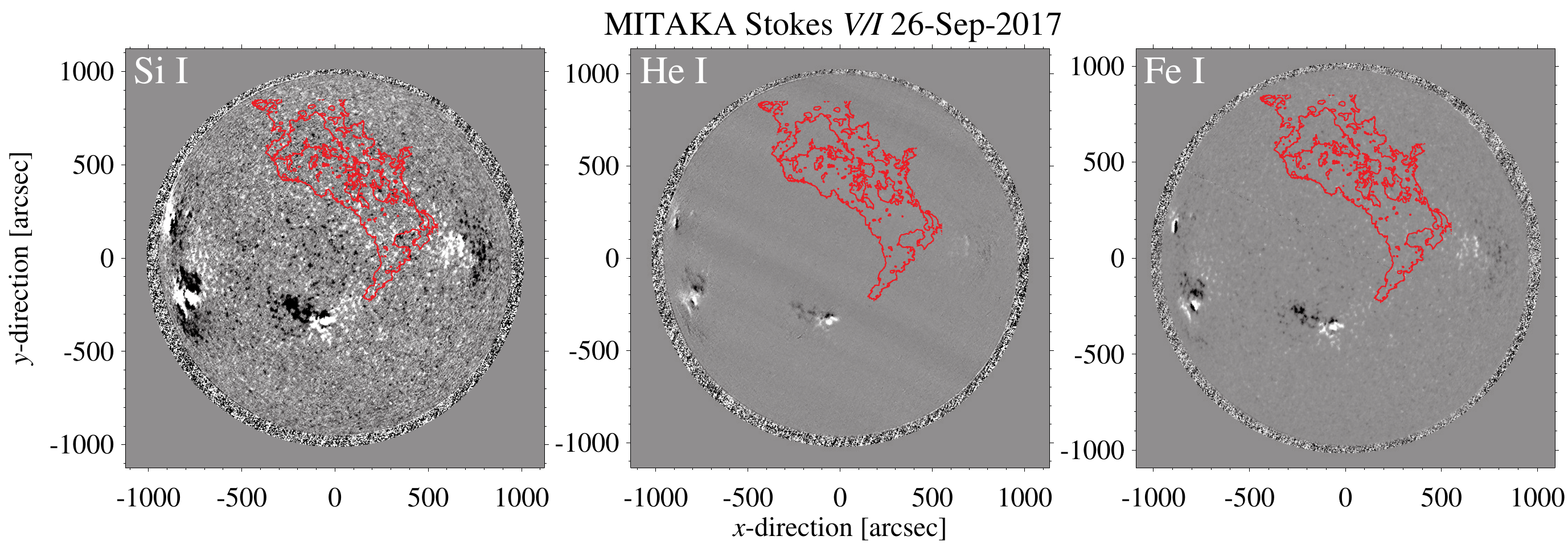}
\caption{Mitaka full-disk Stokes-$V/I$ magnetograms on 26~September 2017 based on the Si\,\textsc{i} 10827\,\AA, \mbox{He\,\textsc{i}}~10830\,\AA, and \mbox{Fe\,\textsc{i}}~15658\,\AA\ spectral lines (\textit{from left to right}). The red contours outline the coronal hole area.}
\label{mitaka_stokesv}
\end{figure}

Since coronal holes appear in chromospheric \mbox{He\,\textsc{i}} 10830\,\AA\ line-core images as regions brighter than the average quiet-Sun, we use a ChroTel filtergram observed at 09:27\,UT on 27~September 2017 to show how the coronal hole appears in the chromosphere (see right panel of Figure~\ref{context_aia_chrotel}). On this day, we encountered seeing conditions which were not optimal during the observing campaign. For this reason, we present in Figure~\ref{mitaka_stokesv} near-infrared Stokes-$V/I$ full-disk magnetograms obtained on 26~September 2017 with SFT at Mitaka, which are based on the \mbox{He\,\textsc{i}} 10830\,\AA, \mbox{Si\,\textsc{i}} 10827\,\AA, and \mbox{Fe\,\textsc{i}} 15658\,\AA\ spectral lines.

\subsection{Coronal Hole Area and Magnetic Field} \label{ch_mag_area}

\begin{figure}
\centerline{\includegraphics[width=0.50\textwidth,clip=]{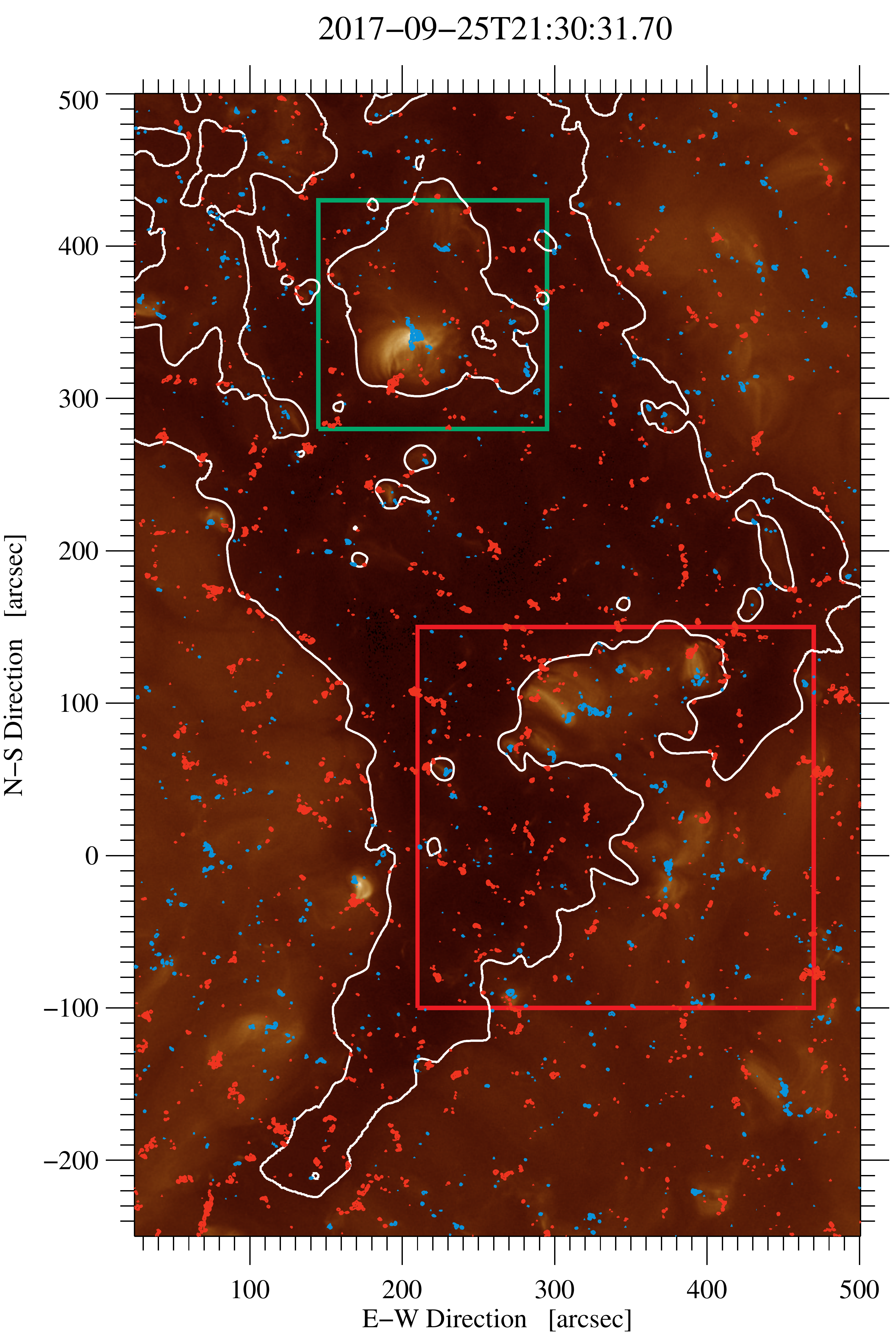}
\includegraphics[width=0.50\textwidth,clip=]{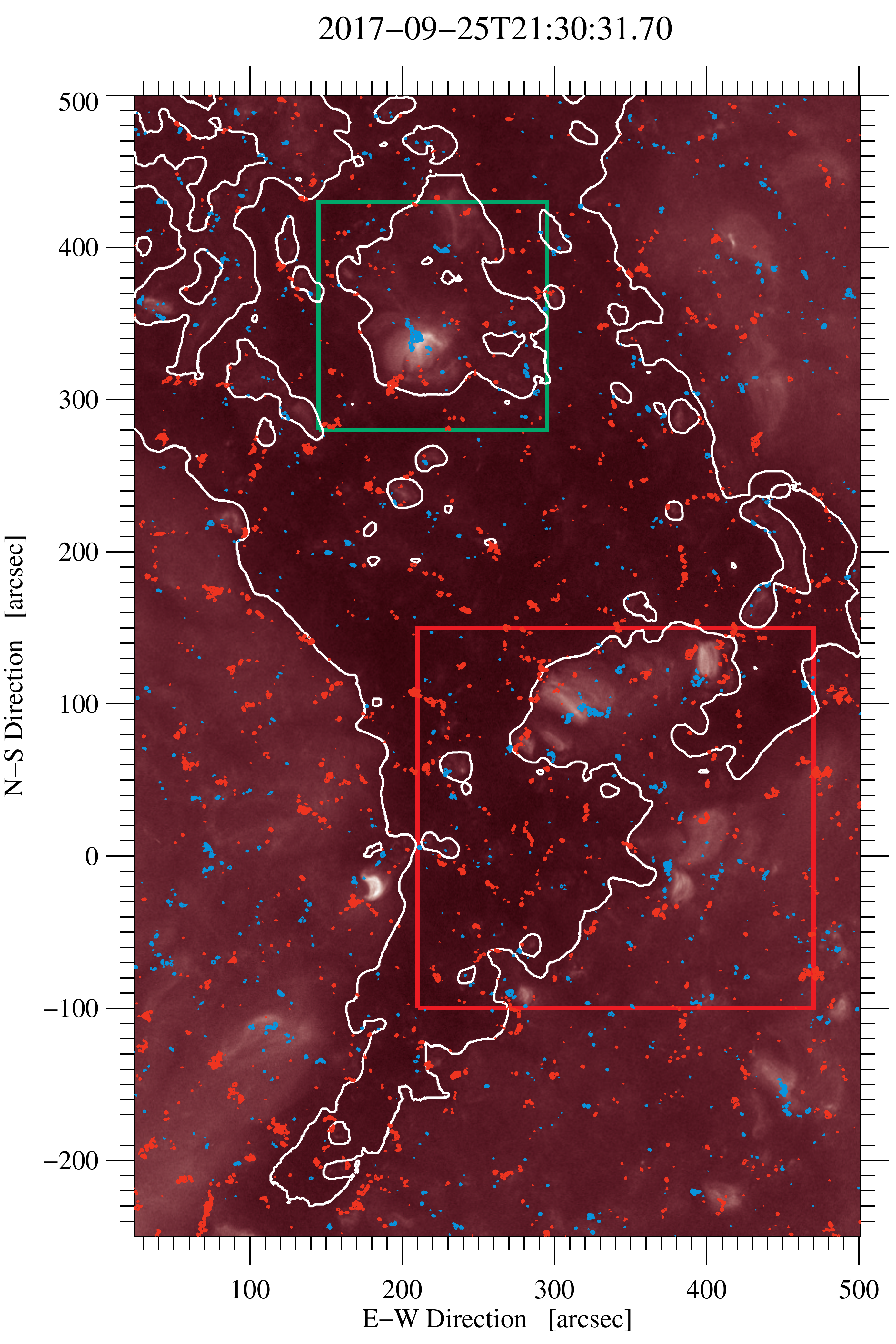}}
\caption{The coronal hole boundary is outlined by white contours in the AIA EUV 193\,\AA\ (\textit{left}) and in 211\,\AA\ (\textit{right}) images. Positive and negative polarities derived from HMI magnetograms are displayed in red and blue, respectively. Contours correspond to the values [45, 100, 200, 300, 500, 700] G for both polarities.}

\label{ch_193_211}
\end{figure}

The coronal hole depicted in Figure~\ref{ch_193_211} is a detailed close-up of the left panel of Figure~\ref{context_aia_chrotel} showing AIA EUV 193\,\AA\ and 211\,\AA\ images. The white contours outline the coronal hole boundaries. These contours are computed by smoothing the coronal hole area with a 15$\times$15~pixel boxcar average. Small-scale magnetic field elements are plotted in red for positive polarity and blue for negative polarity. The green and red boxes refer to ROI\,1 and ROI\,2, respectively, as before. 

The coronal hole area was computed using AIA 193\,\AA\ images. We use a sliding window of $30^{\circ}$ in the $x$-coordinate for tracking the whole coronal hole, from roughly $0^{\circ}$ (disk centre) to $35^{\circ}$ (about three days later). Segmentation and area computation follows also \citet{Hofmeister2017} and \citet{Heinemann2018a}, with similar criteria here, namely correcting the center-to-limb intensity variation and applying a threshold, which is proportional to the median of the entire solar disk. No rebinning or other corrections are performed.
 
A similar procedure as in \citet{Kuckein2012} is applied for magnetic flux computations within the window enclosing the coronal hole. The conservative threshold of $\pm45$\,G for the magnetic flux corresponds to roughly five times the maximum noise level in HMI magnetograms \citep{Liu2012}. Positive and negative fluxes are computed separately, and the magnetic field density was corrected for $\mu = \cos\theta$, where $\theta$ is the heliocentric angle. Values were not corrected for foreshortening. In Figure~\ref{ch_193_211}, red contours correspond to a range of values for the positive polarity elements, and blue contours mark the negative polarity features.

\begin{figure}
\centerline{\includegraphics[width=\textwidth,clip=]{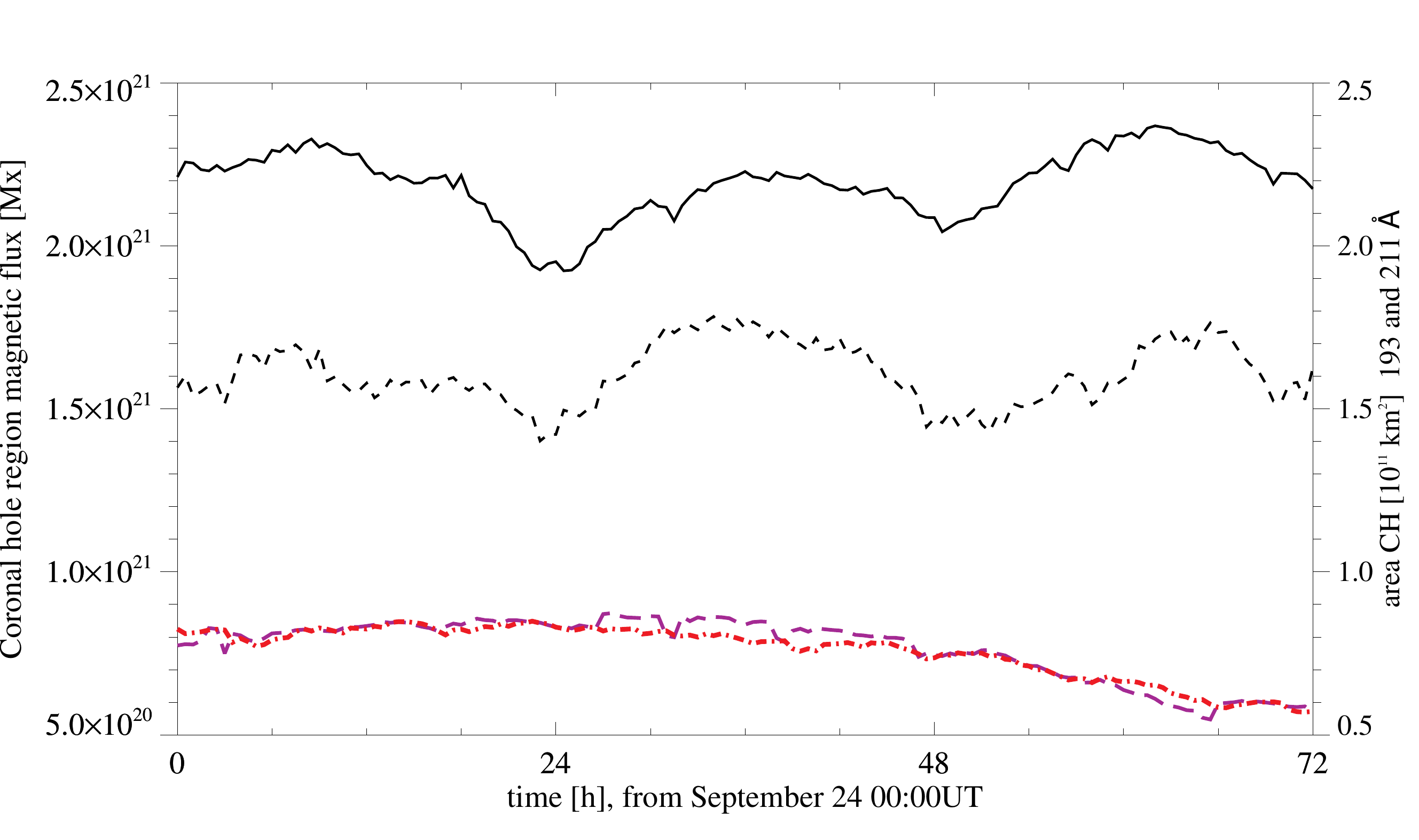}}
    \caption{Temporal evolution of the positive (\textit{solid}) and negative (\textit{short dashed}) fluxes and coronal hole area based on the
    AIA EUV 193\,\AA\ (\textit{red dash-dotted}) and 211\,\AA\ (\textit{purple long-dashed}) images. For display purposes, the negative flux is multiplied by two.}

\label{flux_area_wholeCH}
\end{figure}

The evolution of area with time is displayed in Figure~\ref{flux_area_wholeCH} for AIA EUV 193\,\AA\ and 211\,\AA\ time-series covering a period of 72~hours, from 24 September at 00:00 UT on, which it will be referred as $t_{0} = $0 h in the corresponding text and plot. The areas are expressed in square kilometres. The area evolution on 193\,\AA\ is roughly flat and then steadily decreasing.  The profile in 211\,\AA\ is more irregular than that of 193\,\AA. At around $t_{0} =$ 27~h, after the first flux emergence event happened, the EUV 211\,\AA\ area (in purple) the area steadily decreases, following the decreasing trend of 193\,\AA\ measured area, (in red). The evolution of the positive and negative fluxes inside the tracking window are also displayed. The two broad minima registered around $t_{0} =$ 24 h and after $t_{0} =$ 50 h respectively, correspond roughly to the periods of the investigated magnetic flux emergence and decay timing studied in ROI\,1 and ROI\,2 respectively. The magnetic flux emergence events observed in these two ROIs correspond to a steady decrease of the area of the coronal hole measured in AIA EUV 193\,\AA\ images.

\section{Flux Emergence Events Inside the Coronal Hole}\label{ROIs_sec}

To study the flux emergence events happening within the coronal hole, we divide the coronal hole into two regions. The ROI\,1 is defined as the northern part with a bounding box at coordinates $x=[-50\arcsec, +100\arcsec]$ and $y=[+275\arcsec, +425\arcsec]$ at the reference time 20:00\,UT on 24~September 2017, referred as ($t_{1}= 0$ h) in the text and corresponding plot. The analysed period for the ROI\,1 goes from 20:00\,UT on 24~September 2017 to 02:00\,UT on 26~September 2017, \textit{i.e.} 30 hours. This region ROI\,1 is enclosed with a green box in the left panel of Figure~\ref{context_aia_chrotel}, and in both panels of Figure~\ref{ch_193_211}. 

ROI\,2 is defined as the southern area with a bounding box at coordinates $x=[-110\arcsec$, $+150\arcsec]$ and $y=[-100\arcsec, +150\arcsec]$ at the reference time 10:00\,UT on 24 September 2017, referred as ($t_{2}= 0$ h) in the text and corresponding  plot. The analysed period for ROI\,2 goes from 10:00\,UT on 24~September 2017 to 23:30\,UT on 26 Sep\-tem\-ber 2017, \textit{i.e.} $\sim 62$ hours. This region ROI\,2 is enclosed with a red box in the left panel of Figure~\ref{context_aia_chrotel} and again in both panels of Figure~\ref{ch_193_211}. 

Similarly to the area evolution of the whole coronal hole, we measure the area occupied by the coronal hole in ROI\,1 and ROI\,2. In this case, we do not use a sliding window, since the areas are small. We measure positive, negative, signed (net), and unsigned (total) fluxes in ROI\,1 and ROI\,2. Since positive and negative fluxes exhibit similar temporal profiles, the signed and unsigned fluxes follow the same trends.
 
\begin{figure}
\centerline{\includegraphics[width=\textwidth,clip=]{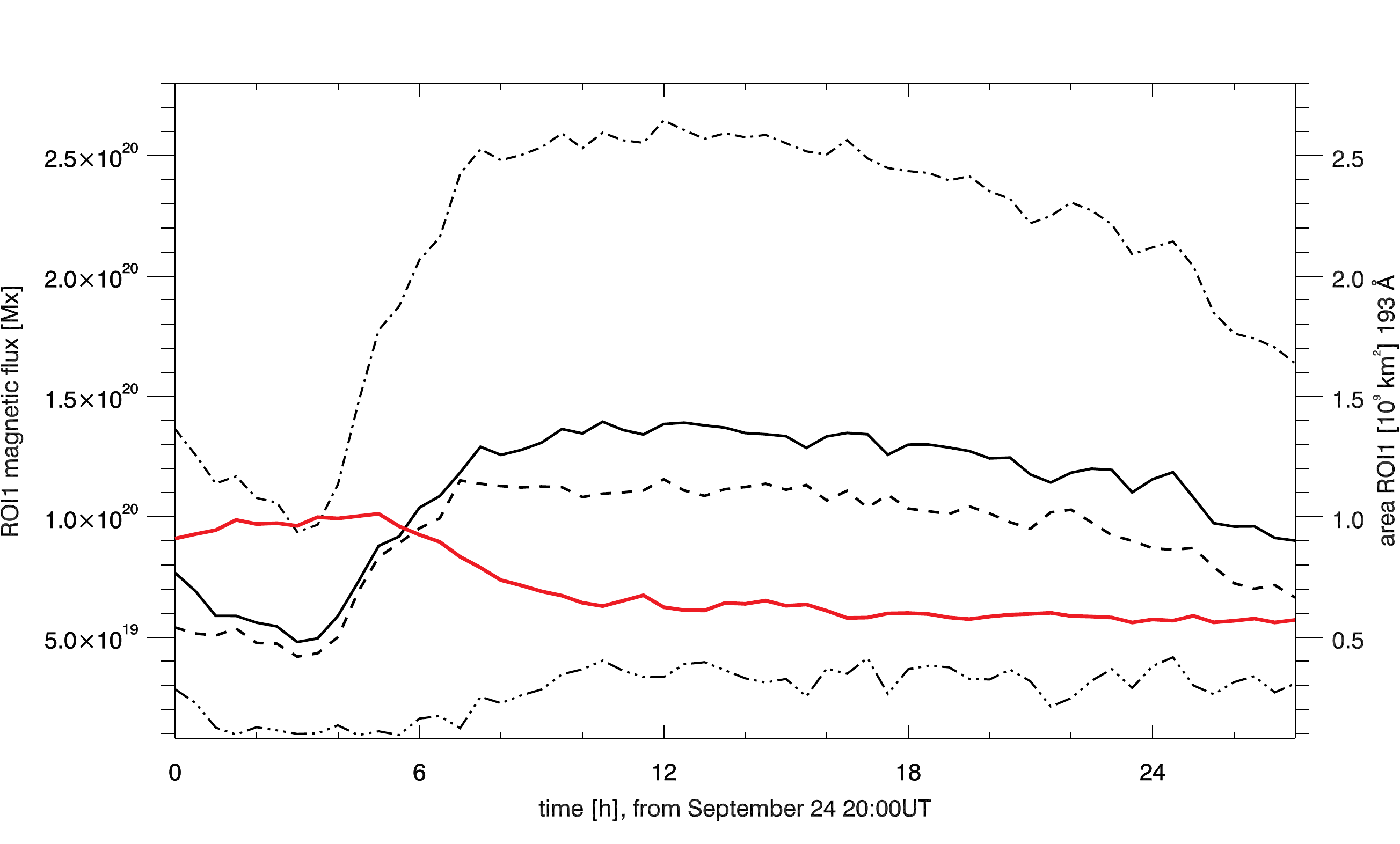}}
\caption{Temporal evolution of the positive (\textit{solid}), negative  (\textit{dashed}), unsigned (\textit{dot-dashed}), and signed (\textit{three-dot-dashed}) fluxes in region ROI\,1. The area measurements (\textit{red}) are based on AIA EUV 193\,\AA\ images.}

   \label{fluxarea_roi1}
\end{figure}

The positive flux is always larger than the negative in both ROI\,1 and ROI\,2. In ROI\,1, the flux emergence event is quite prominent. The unbalanced flux is always positive. The flux increases by a factor of two in ROI\,1, as shown in  Figure~\ref{fluxarea_roi1}. The flux emergence at  ROI\,1 starts at around $ t_{1} \sim 3$ h after the reference time; while the coronal hole area enclosed at ROI\,1 starts decreasing at around $ t_{1} \sim 5$ h.
The decrease from the start to the end of the ROI\,1 series is roughly 30\%. The flux emergence lasts about 4.5~hours from onset (at $t_{1}= 3$ h) to maximum ($t_{1}= 7.5$ h) with an emergence rate of $(6.3\pm 0.3) \times 10^{17}$\,Mx\,min$^{-1}$ for the unsigned flux. The emergence rate is $(3.3 \pm 0.1) \times 10^{17}$\,Mx\,min$^{-1}$ for the positive flux.

\begin{figure}
\centerline{\includegraphics[width=\textwidth,clip=]{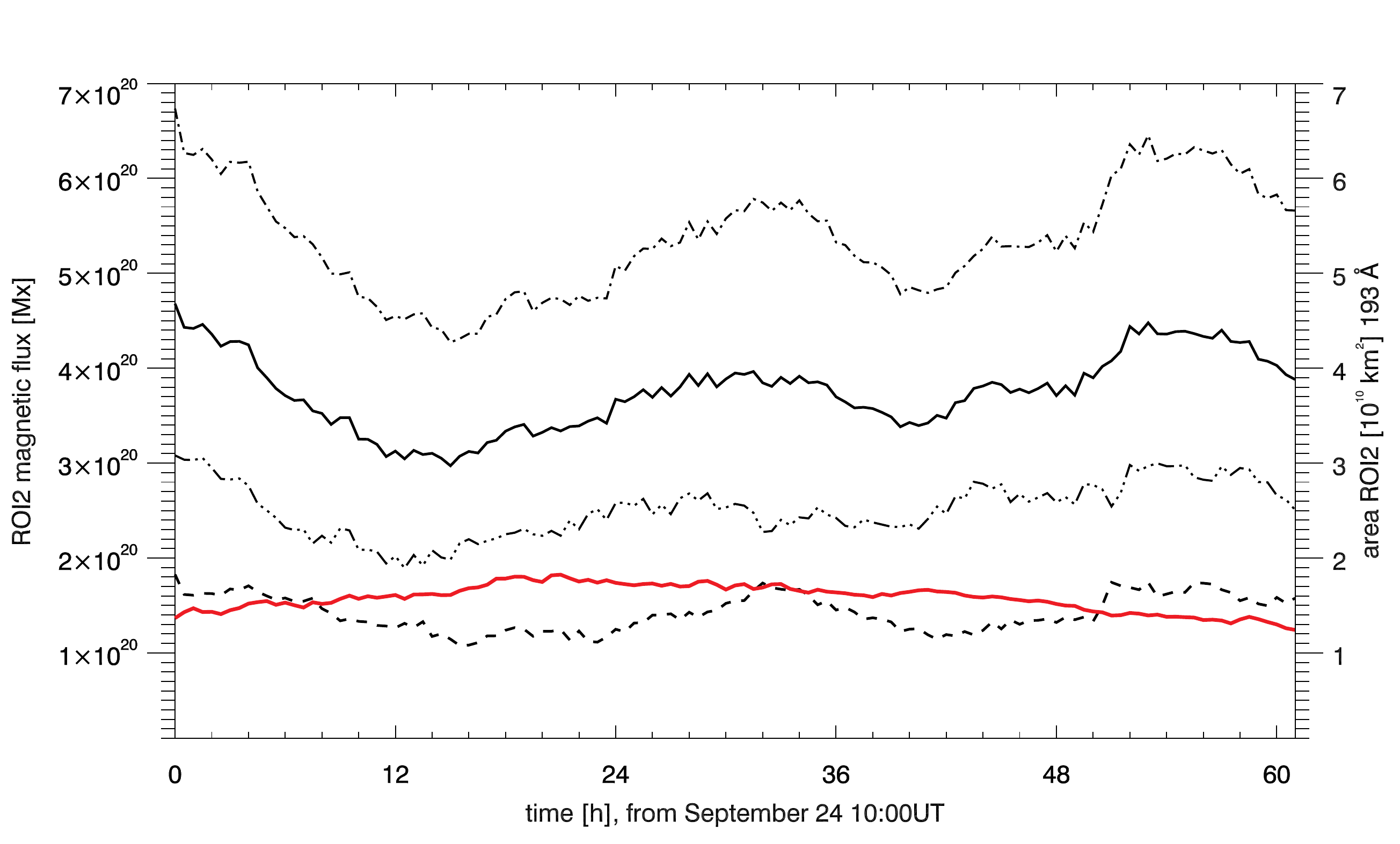}}
\caption{Temporal evolution of the positive (\textit{solid}), negative  (\textit{dashed}), unsigned (\textit{dot-dashed}), and signed (\textit{three-dot-dashed}) fluxes in region ROI\,2. The area measurements (\textit{red}) are based on AIA EUV 193\,\AA\ images.}

\label{fluxarea_roi2}
\end{figure}

We study the flux emergence in ROI\,2 shown in Figure~\ref{fluxarea_roi2}. In ROI\,2, the flux emergence is recurrent from about 01:00\,UT on 25~September 2017 ($t_{2}=15 $ h) to the next day. At that time, another flux emergence event starts at about 00:00\,UT on 26~September 2017 ($t_{2}=38 $ h), which overlaps with another event at about 10:00\,UT. We compute the emergence rate for the last two coupled events during a 13.5-hour time period (from $t_{2} = 38$\,h to $t_{2} = 51.5$\,h, meaning from 26 September 2017 at 00:00\,UT to 26 September 2017 at 13:30\,UT). This emergence rate yields $(1.8 \pm 0.1) \times 10^{17}$\,Mx\,min$^{-1}$ for the unsigned flux and $(1.14 \pm 0.01) \times 10^{17}$\,Mx\,min$^{-1}$ for the positive flux.

\begin{figure}      
\includegraphics[width=0.5275\textwidth,clip=]{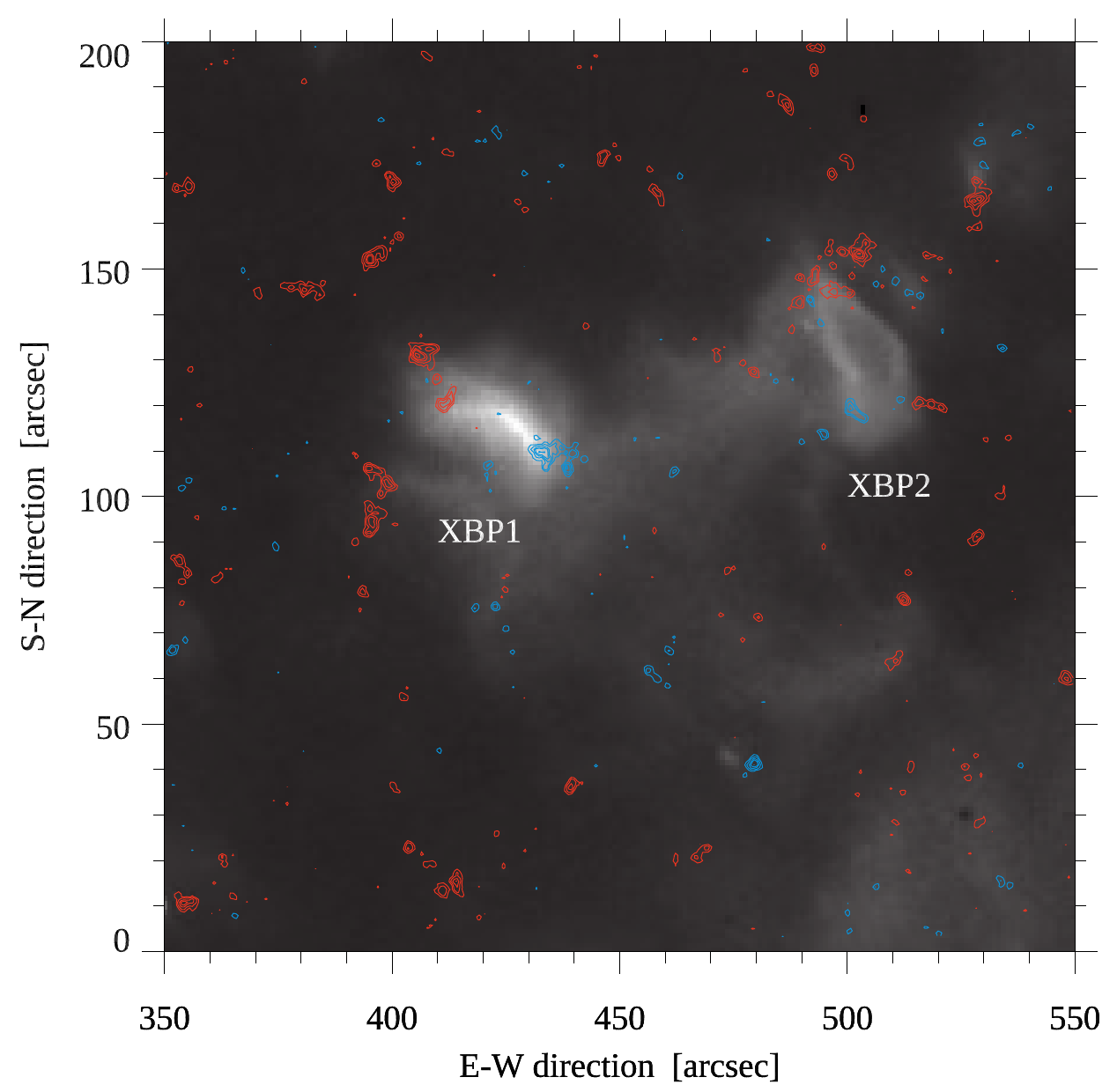}
\hspace{-2mm}
\includegraphics[width=0.4715\textwidth,clip=]{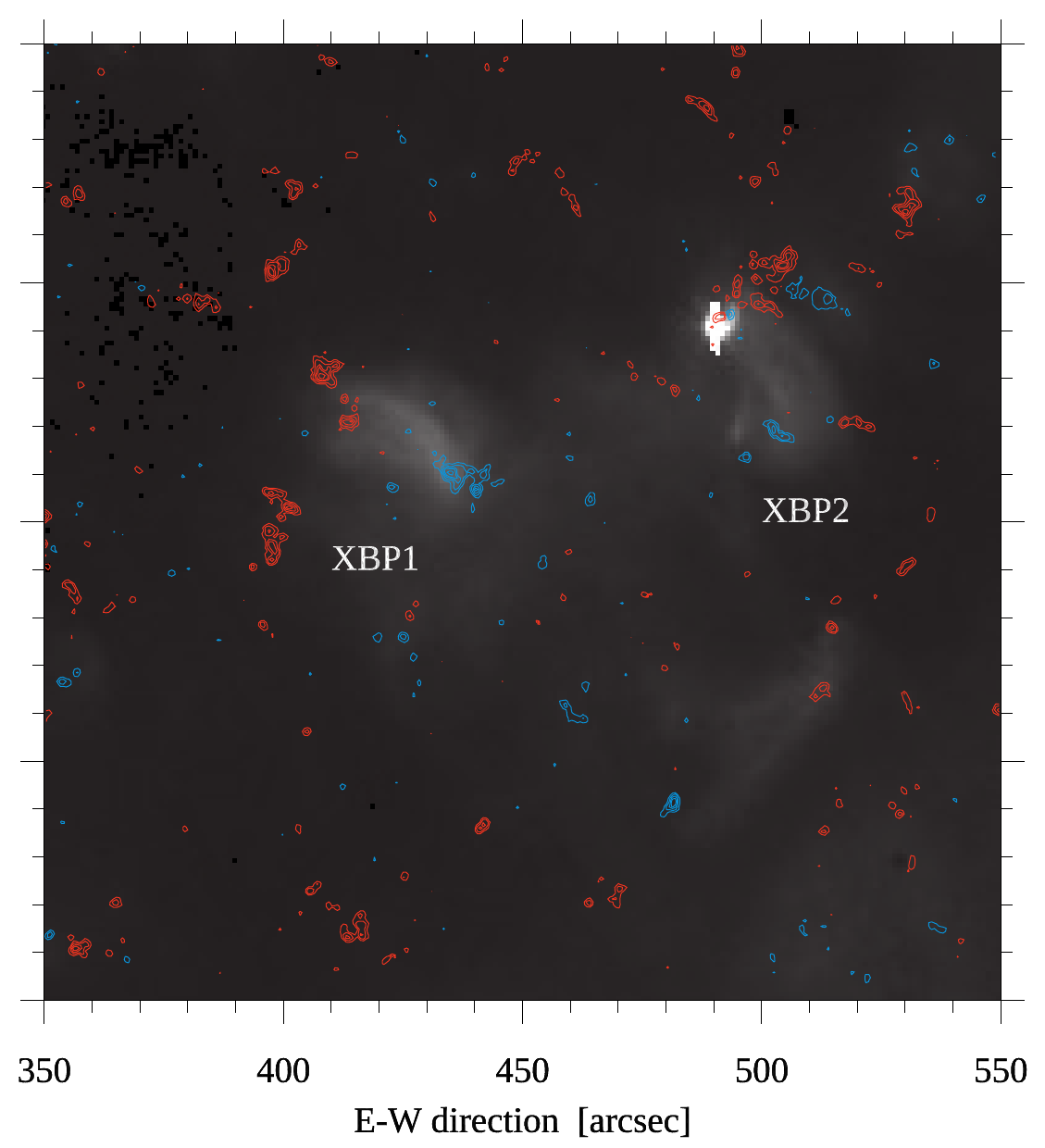}
\caption{Hinode/XRT images of ROI\,2 taken on 26 September 2017 at  11:15\,UT ({\it left}) and 11:31\,UT ({\it right})
showing two X-ray Bright Points XBP\,1 and XBP\,2. The red and blue contours mark positive and negative flux, respectively.}
\label{fig_xbps}
\end{figure}

These two emergence events are related to flux emergence in already formed XBPs in  ROI\,2. Here we show brightenings that are likely related to reconnection of the emerged flux with the ambient field. The scale of the occurrence is supergranular. This complementary picture of magnetic activity and heating process ROI\,2  is shown in Figure~\ref{fig_xbps}. Two Hinode/XRT images taken on 26 September 2017 about 16\,min apart exemplify two X-ray bright points XBP\,1 and XBP\,2 occurring in ROI\,2. Both XBPs have loop-like structure bridging the opposite polarities. The north footpoint of XBP2 at $(x,y) = (500\arcsec, 150\arcsec)$ features a mixed polarity leading to intense brightening at 11:31\,UT (right panel of Figure~\ref{fig_xbps}). 

The total area of the coronal hole enclosed in ROI\,2 shrinks steadily by about 40\% at this phase from maximum area until the end of the time series shown in Figure~\ref{fluxarea_roi2}. The maximum area is reached at around $t_{2} \sim 19$ hours. Interestingly, the first phase (from $t_{2}= 0$ h to $t_{2} \sim 19$ hours) depicts a phase of cancellation, where the area of the coronal hole at ROI\,2 grows by about 30\% of the initial size. This phase is analysed in more detail in Section~\ref{fluxcancel_subsec}.

 As summary for this Section, the flux emergence profiles of ROI\,1 and ROI\,2 are quite different, exhibiting a single fast emergence event in the interior of the CH, and recurrent slowly evolving events at the boundary of the coronal hole, respectively.

\section{Coronal Hole Area Reduction and Growth due to Flux Emergence and Cancellation}\label{emergcancel_sec}

In this section, we provide evidence that flux emergence and coronal hole area reduction are related, whereas reconnection events lead to an increase of the coronal hole area.

\subsection{Flux Emergence Event at ROI 1}

 Here we investigate a flux emergence event occurring in ROI\,1. As in the previous images, the positive and negative magnetic fluxes are shown in red and blue contours, respectively, for the strong magnetic field structures observed by SDO/HMI. Their background corresponds to thresholded AIA images with the outlined boundary of the coronal hole. A circle is chosen for a finer selection of small structures within the ROIs. The flux is computed inside these circular regions.

\begin{figure}
\includegraphics[width=1.0\textwidth,clip=]{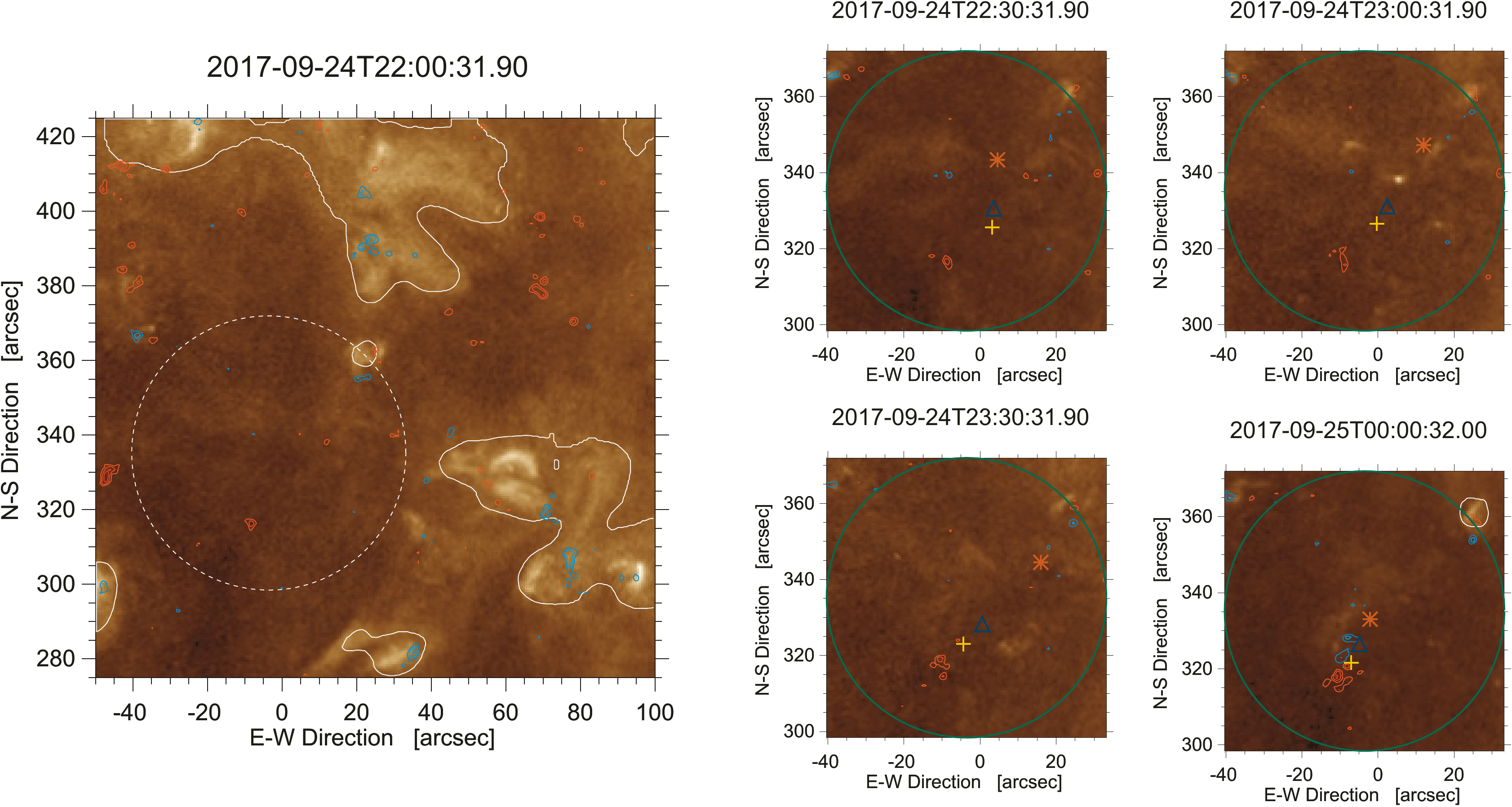}
\caption{The flux emergence site in ROI\,1 (\textit{left}) and four panels depicting its temporal evolution (\textit{right}), zooming in on details before the flux emergence (Phase~\textsc{0}). Positive and negative polarities are displayed in red and blue, respectively. The yellow plus sign corresponds to the barycentre of positive flux elements, the red asterisk to negative flux elements, and the blue triangle to the unsigned magnetic flux barycentre.}
   \label{flux_em_phase0}
   \end{figure}

 Figure~\ref{flux_em_phase0} shows ROI\,1 some hours before the flux emergence event occurs. In the left panel, we display the general location of the flux emergence marked by the dashed circle. In the smaller panels at right we show the details within the circle with a cadence of 30~minutes. Three strong magnetic flux concentration systems, where the magnitude of the magnetic field is larger than 45\,G, were found within the circular FOV. In the following plots, the yellow plus sign corresponds to the barycentre of positive magnetic elements, the red asterisk to negative flux elements, and the blue triangle is related to the unsigned magnetic flux barycentre. Before the onset of the flux emergence, the location is magnetically quiet and dark with only sporadic small-scale magnetic elements appearing and fading soon, which do not affect to the average intensity of the coronal hole. However, we can already observe in the surroundings that magnetic bipoles lead generally to brighter, often loop-like structures limiting the coronal hole and decreasing its size. 

\begin{figure}
\includegraphics[width=1.0\textwidth,clip=]{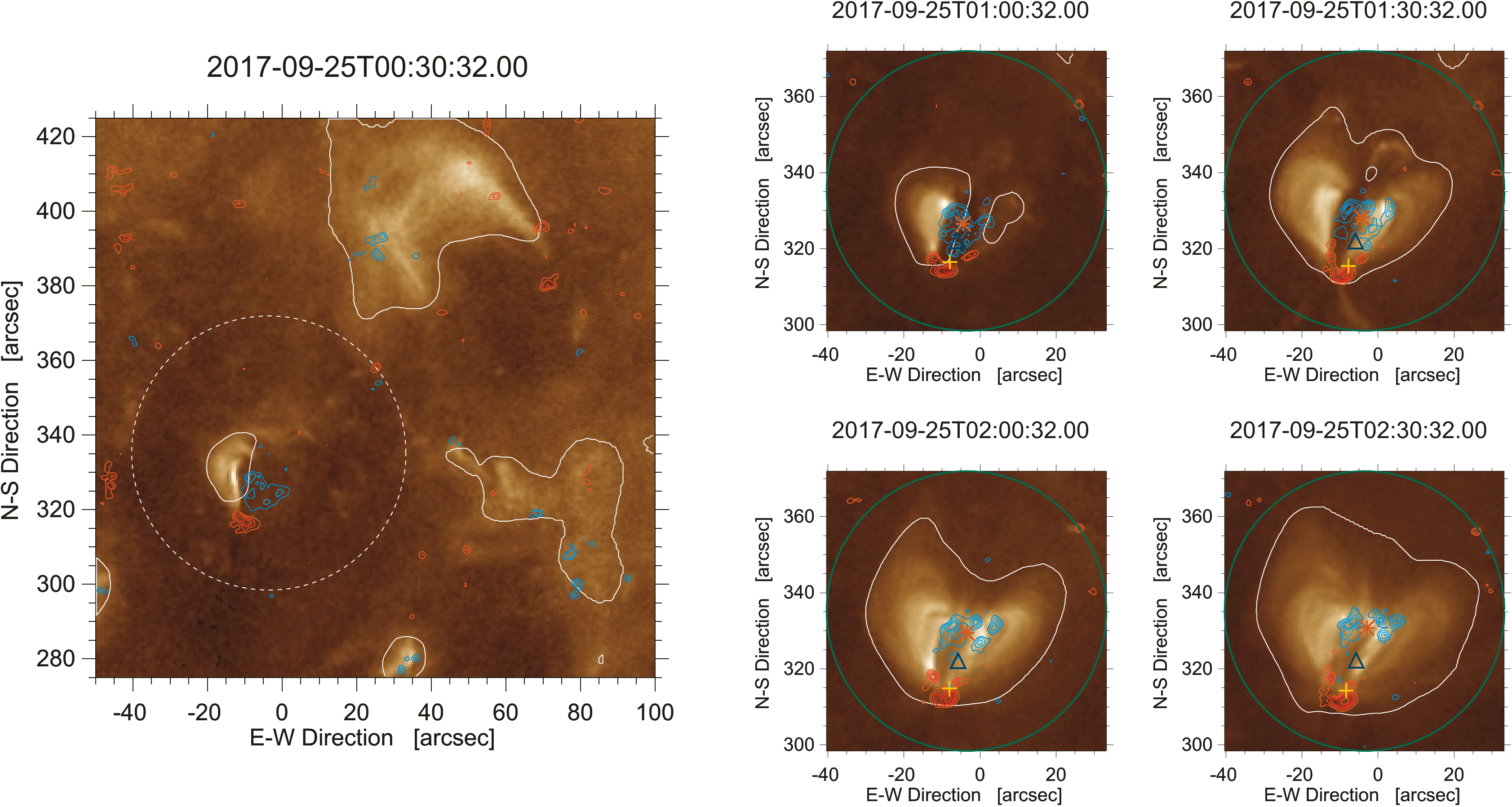}
\caption{Temporal evolution of ROI\,1, zooming in on details of the initial phase of flux emergence (Phase~\textsc{I}). The white contours outline bright EUV intensity structures which exceed the value of the coronal hole threshold and do not count as the coronal hole area. The red and blue contours mark positive and negative flux, respectively. The contour lines refer to 45\,G,~100\,G,~200\,G,~300\,G,~500\,G, to even 700\,G for both polarities.}
\label{flux_em_phase1}
\end{figure}

 At the moment of flux emergence (Phase~I), shown in Figure~\ref{flux_em_phase1}, the coronal area starts to brighten up at the location where the new flux emerges. The new flux appears as a diffuse bipolar field, as shown with blue and red contours in the left panel of Figure~\ref{flux_em_phase1}). To the left of the flux emergence site, a bright loop structure forms and reduces the area of the coronal hole. In the subsequent two hours, as illustrated in the four small panels, more magnetic field emerges and coalesces in two footpoints connected by a coronal loop structure. The brighter and therefore heated area increases in size and further shrinks the area considered as coronal hole.

\begin{figure}       
\includegraphics[width=1.0\textwidth,clip=]{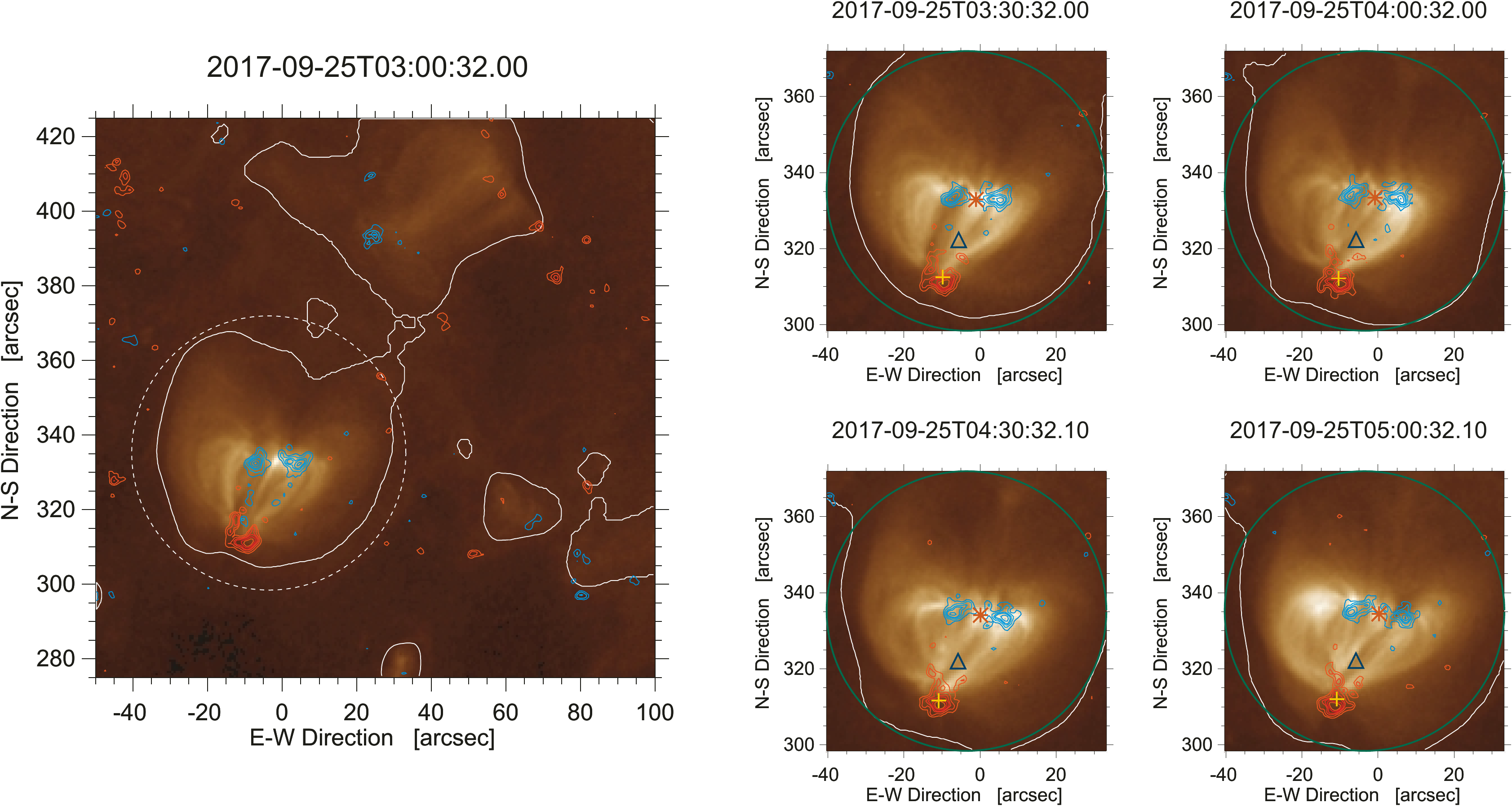}
\caption{Same as Figure~\ref{flux_em_phase1} but for the phase of stabilization and maximum spatial extent (Phase~\textsc{II}).}
\label{flux_em_phase2}
\end{figure}

 In the next hours, we notice a stabilization and only slow growth, as displayed in  Figure~\ref{flux_em_phase2}. However, adjacent areas, which are probably slightly heated due to their own magnetic field activity, get slowly incorporated in the brightened area at coordinates [+30\arcsec,\, +360\arcsec]. Furthermore, several other bipolar magnetic regions form bright areas, which reduces the local area of the coronal hole (marked as white contours in Figure~\ref{flux_em_phase2}).

\begin{figure}       
   \includegraphics[width=1.0\textwidth,clip=]{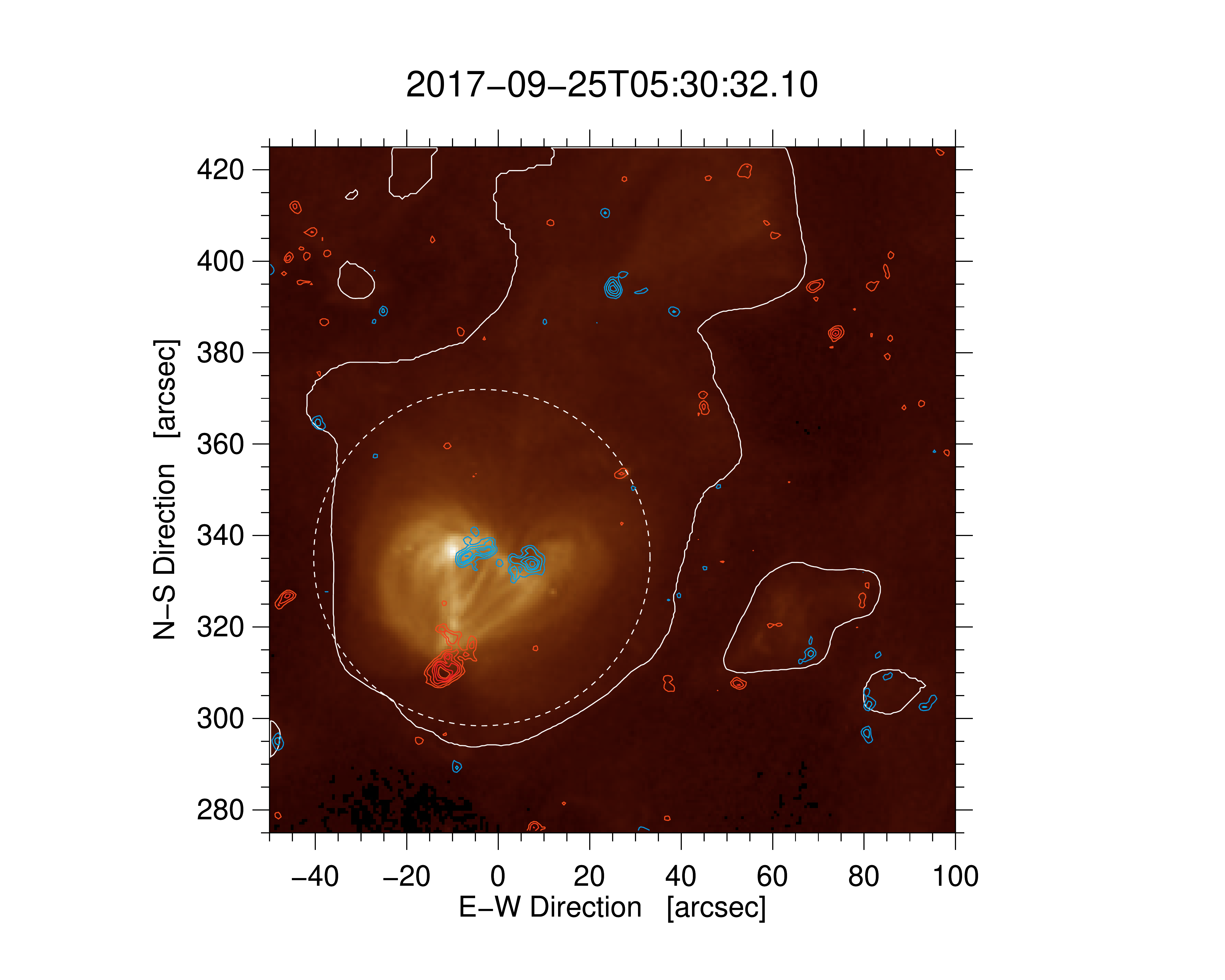}
              \caption{The stable phase with further heating and merging to adjacent magnetically heated areas (Phase~\textsc{III}).}
   \label{flux_em_phase3}
   \end{figure}

 In the last phase of the emergence, displayed in Figure~\ref{flux_em_phase3}, the heated and brightened area stabilizes with respect to the quiet coronal hole but still extends slightly to the neighbouring area and thus some parts start to merge with other newly brightened regions, \textit{e.g.} to the northeast in Figure~\ref{flux_em_phase3}, where other magnetic bipolar activity led to brightened areas.
 
\begin{figure}       
   \includegraphics[width=1.0\textwidth,clip=]{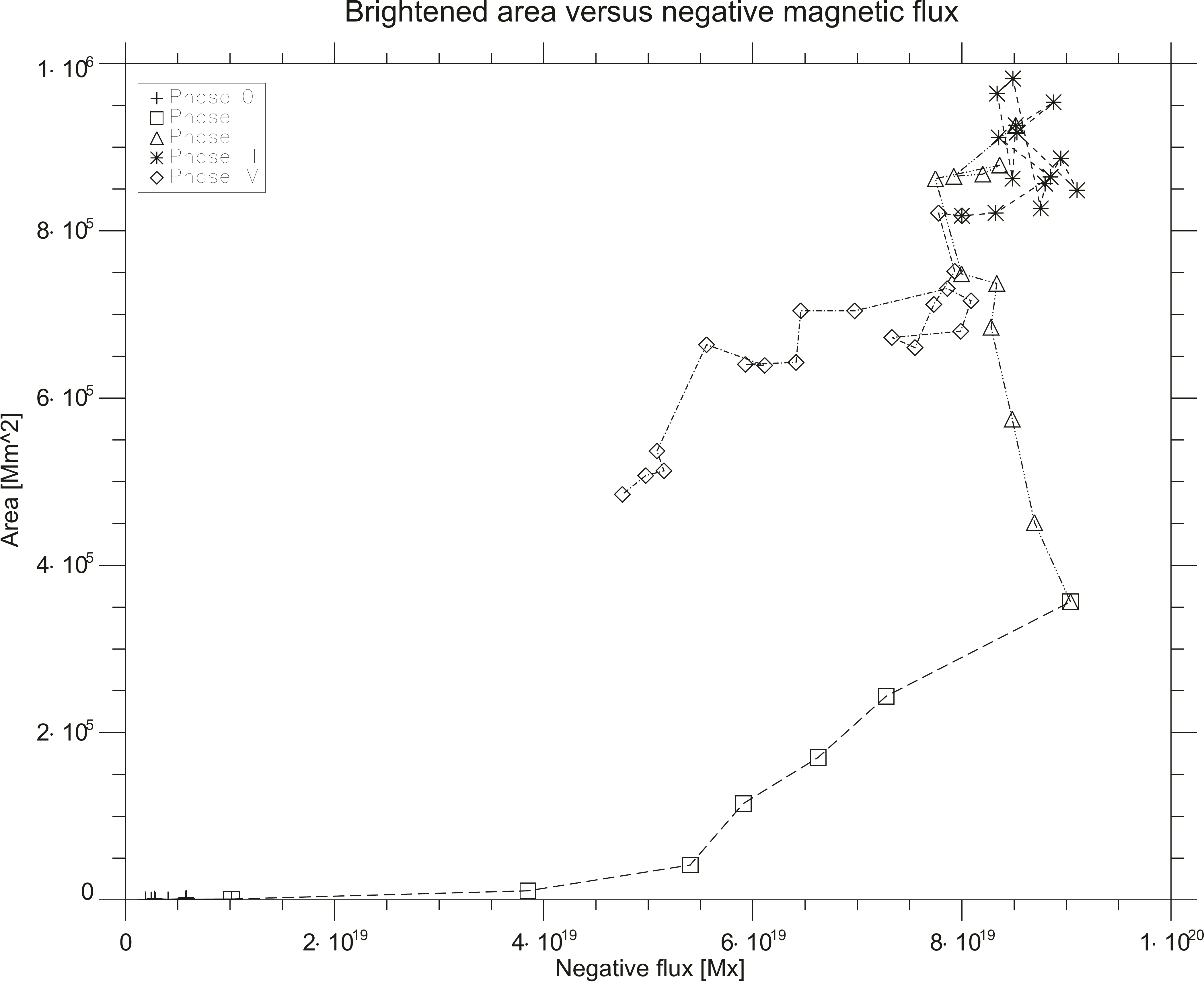}
              \caption{The size/negative flux relationship evolution is shown. The curve runs from the lower left corner (0 flux/0 area) and Phase~\textsc{0}, over the upper right corner (Phase~\textsc{III}; stable phase), to an end point in the centre of the figure. Fluxes are plotted in absolute value.}
   \label{flux_em_evol}

   \end{figure}

 The described and inspected relationship is also shown in Figure~\ref{flux_em_evol}. The negative flux element is chosen for the representation; however, the positive may be equivalent, since structure is bipolar with similar fluxes. Here we show the temporal evolution of the brightened area versus negative flux relationship, in absolute value. The whole curve is split in five phases: Phase~\textsc{0} is prior to the flux emergence and shown as a concentration of measurement points, marked by crosses scattered close to the [0, 0] coordinates of the diagram. In Phase~\textsc{I}, the flux emergence event increases dramatically the negative polarity magnetic flux while also slowly increases the size of the brightened area. In the next Phase~\textsc{II}, the footpoints of the newly formed magnetic bipolar island start to separate. Thus the area of the newly brightened emergence region is increasing while the negative flux strength is only very slightly dropping. In Phase~\textsc{III}, the region stabilizes and exists quite unaltered for about 20 measurement points or ten hours while finally in Phase~\textsc{IV}, the region starts slowly to disappear over periods of several hours.   
   
  Thus we conclude that a typical reduction process in area of coronal holes can be formation of islands of newly brightened coronal hole area by flux emergence events. After a quiet time (Phase~\textsc{0}), a rapidly evolving flux emergence event  occurs on timescales of fraction of hours to hours (Phase~\textsc{I}) leading to the formation of bipolar magnetic regions with heated loop structures in between reducing the dark area of the coronal hole. After a gradual growth (Phase~\textsc{II}), the newly brightened area reaches its maximum after a few hours and depending on the surrounding magnetic field configuration merges with other newly brightened areas (Phase~\textsc{III}).
  
\subsection{Horizontal Flow Fields}   

\begin{figure}[t]    
\includegraphics[width=1.0\textwidth]{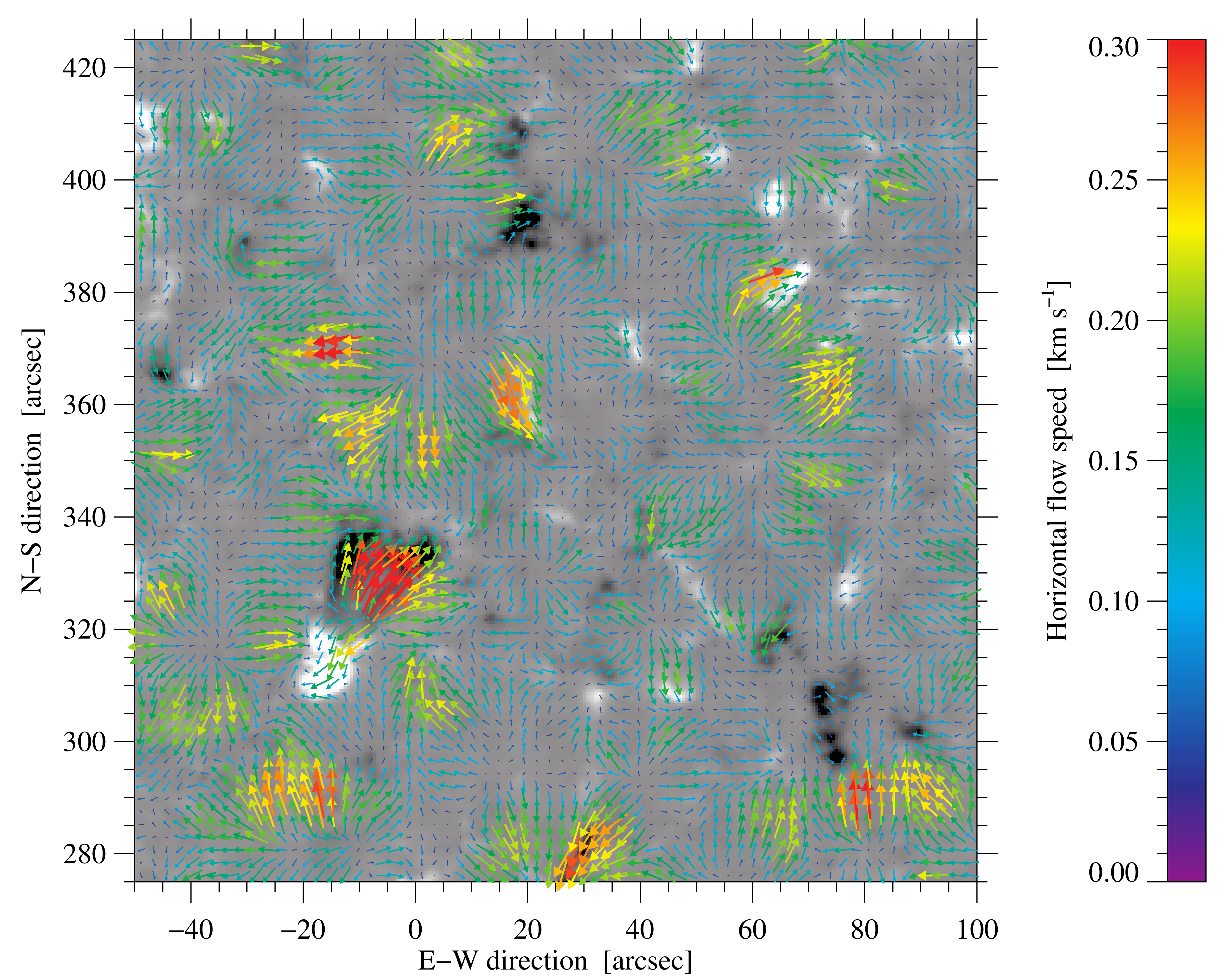}
\caption{Horizontal flow speed derived with DAVE for ROI\,1, which contains a quiet-Sun region and a small EFR. The vectors represent magnitude and direction of the persistent flux transport velocities, and the quantitative value of the flow speed is given by the rainbow-colored scale bar. The background image is a 10-hour time-averaged magnetogram (20:00\,UT on 24~September 2017 -- 06:00\,UT on 25~September 2017).}
\label{DAVE_ROI1}
\end{figure}

\begin{figure}[t]     
\includegraphics[width=1.0\textwidth]{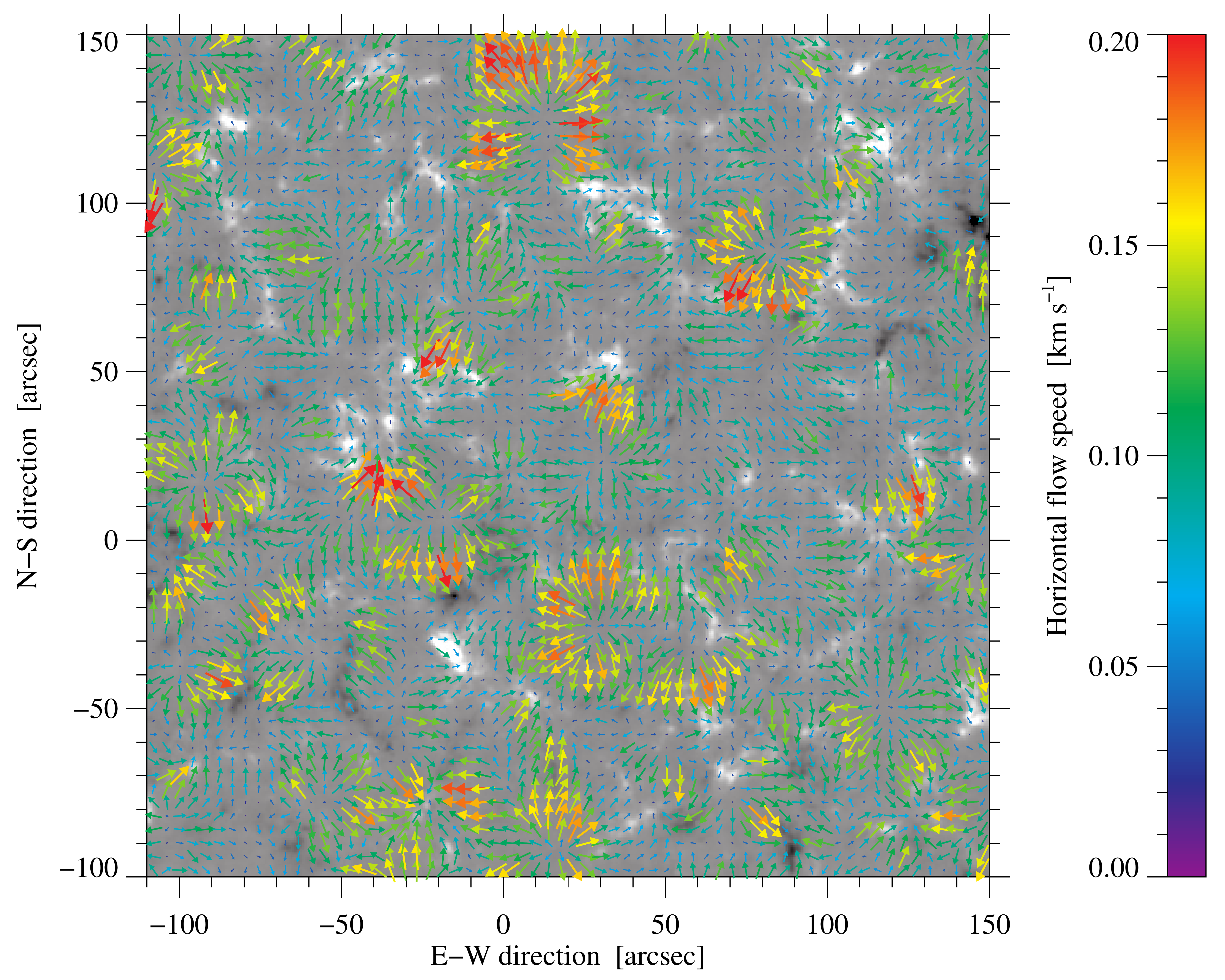}
\caption{Horizontal flow speed derived with DAVE for ROI\,2, which contains quiet-Sun region and a small EFR. The vectors represent magnitude and direction of the persistent flux transport velocities, and the quantitative value of the flow speed is given by the rainbow-colored scale bar. The background image is a 38-hour time-averaged magnetogram (02:00\,UT on 25~September 2017 -- 16:00\,UT on 26~September 2017).}
\label{DAVE_ROI2}
\end{figure}

 The horizontal flux transport velocities were derived for ROI\,1 and ROI\,2 using time-series of SDO/HMI line-of-sight (LOS) magnetograms with a cadence of 45\,s. The full-disk magnetograms were compensated for differential rotation, using the first image in each time-series as a reference. The coordinates of the horizontal flow maps in Figures~\ref{DAVE_ROI1} and~\ref{DAVE_ROI2} refer to these time stamps. The magnetic flux density was corrected for the cosine of the heliocentric angle $\mu = \cos\theta$ assuming that the magnetic field lines are to first order perpendicular to the solar surface in quiet-Sun regions. However, geometric foreshortening was not corrected. One pixel corresponds to roughly 367\,km in the FOV of both ROIs. The duration of both time-series differ significantly, \textit{i.e.} 10~hours from 20:00\,UT on 24~September 2017 to 06:00\,UT on 25~September 2017 and 38~hours from 02:00\,UT on 25~September 2017 to 16:00\,UT on 26~September 2017 for ROI\,1 and ROI\,2, respectively. Since both ROIs contain mainly quiet-Sun regions, a sliding average was applied to the magnetogram series with a temporal window of $\pm$6~minutes to track reliably on small-scale magnetic features. Photon statistics imply an improved signal-to-noise (S/N) ratio of four. However, the real improvement of the S/N-ratio is about a factor of three. The extracted and aligned time-series were subjected to the \textsf{Differential Affine Velocity Estimator} \citep[DAVE,][]{Schuck2005, Schuck2006} to determine the horizontal flux transport velocities. The underlying physics implemented in DAVE includes the magnetic induction equation and an affine velocity profile. Temporal and spatial derivatives of magnetic fields are a prerequisite to compute the horizontal flows. The spatial derivatives were computed by applying the Scharr operator \citep{Scharr2007}, and a five-point stencil with a time step of 15~minutes between neighboring magnetograms was used for temporal derivatives. The use of five-point stencil resulted in the reduction of both time sequences by 60~minutes, \textit{i.e.} 30~minutes in the beginning and in the end. The sampling window employed by DAVE is about 4\,Mm $\times$ 4\,Mm. This implementation of DAVE is similar to the one described in \citet{Verma2016}.

 In evaluating the horizontal flow properties it is important to distinguish between persistent time-averaged and instantaneous flows. The horizontal flow maps in Figures~\ref{DAVE_ROI1} and~\ref{DAVE_ROI2} represent long-term averages, where the global means of the $x$- and $y$-components of the velocity vector were subtracted before taking the average. In addition, the flow vectors represent a spatial average over a neighborhood of 5$\times$5 and 8$\times$8 pixels for ROI\,1 and ROI\,2, respectively. Furthermore, taking the sliding average across the magnetogram series and the size of the DAVE sampling window lead to additional smoothing. The largest instantaneous flow speeds reach up to 3.0\,\kms\ and 2.3\,\kms\ for ROI\,1 and ROI\,2, respectively, whereas the corresponding mean flow speeds are about $0.22 \pm 0.13$\,\kms\ and $0.11 \pm 0.15$\,\kms. Regions within the emerging flux region in ROI\,1, where the magnitude of the flux density is above 100\,G, exhibit the strongest horizontal flows with a mean value of $0.33 \pm 0.22$\,\kms, which is about 50\% higher than the values for the surrounding quiet Sun. These motions mainly result from the separation of the opposite-polarity regions of the EFR. The largest persistent flow speeds reach up to 0.47\,\kms\ and 0.26\,\kms, respectively, whereas the corresponding mean flow speeds are about $0.11 \pm 0.06$ and $0.09 \pm 0.4$\,\kms. The different characteristic flow speed values for the two ROIs result from their dissimilar morphology and the discrepancy of the temporal averages (10 \textit{vs.} 38~hours). The standard deviation in the above context refers to a variation within the ROI, \textit{i.e.}, a physical property of the observed region on the Sun, and not to an error estimate. In general, the parameter choice and preprocessing of the time-series data have a much more significant impact on the flow speed values than the numerical errors.


   \subsection{Flux Cancellation at ROI 2}
   \label{fluxcancel_subsec}
   
\begin{figure}       
   \includegraphics[width=1.0\textwidth,clip=]{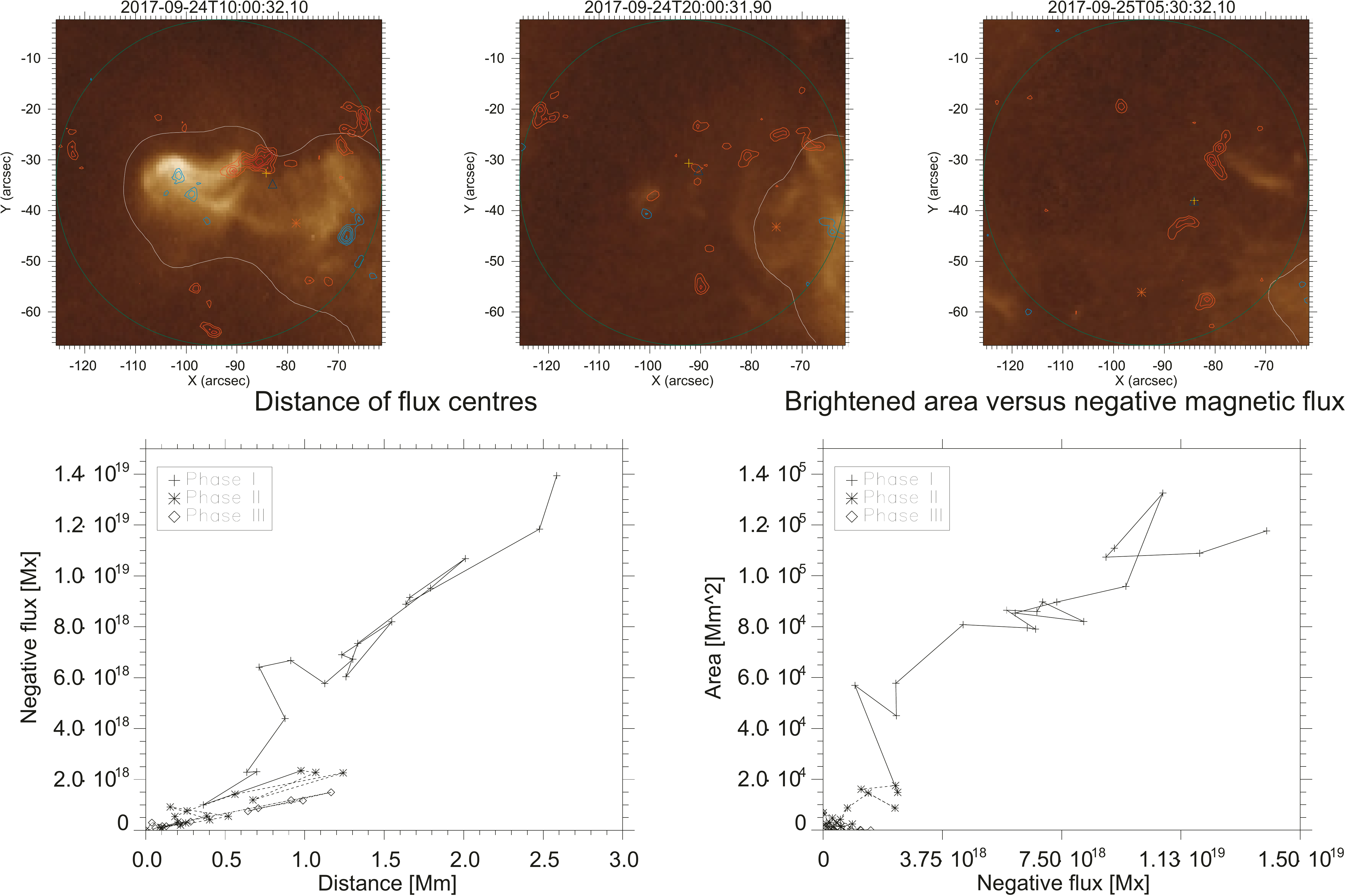}
   \caption{A flux cancellation event leading to dimming of the brightened area and thus to the increase of the coronal hole area. {\it Top:} Evolution of the magnetic field configuration
   in time. The red and blue contours mark positive and negative flux, respectively. The yellow plus sign corresponds to the barycentre of positive flux elements, the red asterisk to negative flux elements, and the blue triangle to the unsigned magnetic flux barycentre. {\it Bottom:} Evolution of the negative total magnetic flux (in absolute value) within the circular area in the top panels versus distance of the barycentres between positive and negative flux ({\it left panel}) and versus area of the brightened region within the coronal hole ({\it right panel}).}
   \label{cancellation}
   \end{figure}

   In this subsection we show that the subsequent flux cancellation leads to shrinking the brightened area within the coronal hole and thus to increase the coronal hole area. The top panels of Figure~\ref{cancellation} show a selected subregion within ROI\,2. The left panel shows strong positive and negative magnetic field elements with a brightened area in between the two polarities of field elements. The magnetic elements get closer slowly, as displayed in the lower left panel of Figure~\ref{cancellation}, which depicts the strength of the negative polarity versus distance between the barycentres of the magnetic field polarities. Time runs from the upper right corner (large distance and large flux) to the lower left corner (small barycentre distance and practically no negative magnetic flux left). Hereby we can identify three phases of evolution. In Phase~\textsc{I}, the magnetic flux decreases with decreasing distance between the two flux centres. This indicates an ongoing magnetic flux cancellation. Phase~\textsc{I} is also depicted in the in the top left panel followed by Phase~\textsc{II} shown in the middle panel. The negative polarity has already disappeared and the atmosphere cools down and becomes dark leading to a growth of the coronal hole area and to a shift of the coronal hole boundary to the lower right corner. There are still some remains of the negative polarity. This explains an increase of the barycentre distance in Phase~\textsc{II}, which is shown in the bottom left panel of Figure~\ref{cancellation}. Finally, in Phase~\textsc{III}, shown at the top right panel, the negative polarity close to the region boundary has vanished and the boundary of the coronal hole moved out of the inspected area yielding the end of the evolutionary curve in the lower left panel.
  
   The lower right panel of Figure~\ref{cancellation} shows the evolution of negative magnetic flux versus size of the brightened area within the coronal hole. Again the curve is running from the top right corner to the lower left corner indicating that the loss of negative polarity within the predominantly positive polarity coronal hole (thus the increase of uni-polarity and open flux) leads to shrinking the brightened area and increasing the coronal hole area.
   
\begin{figure}       
   \includegraphics[width=1.0\textwidth,clip=]{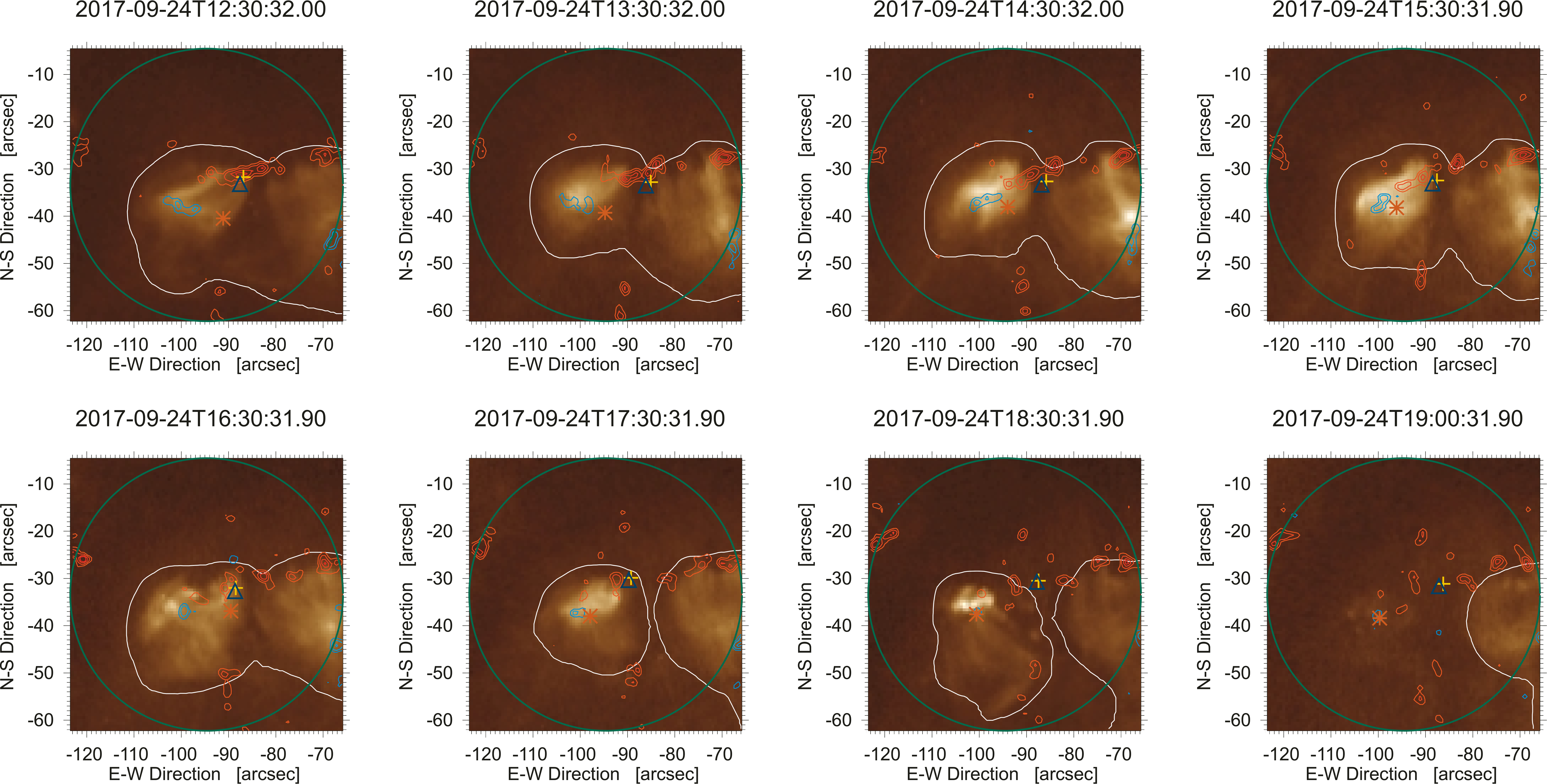}
   \caption{The flux cancellation event from Figure~\ref{cancellation} in more detail. Time runs from the top left to bottom right panel in an one hour cadence sequence. The last panel is taken half an hour apart of the second last one due to rapid evolution. The red and blue contours mark positive and negative flux, respectively. The yellow plus sign corresponds to the barycentre of positive flux elements, the red asterisk to negative flux elements, and the blue triangle to the unsigned magnetic flux barycentre.}
      \label{dis_evol}
   \end{figure}

   Figure~\ref{dis_evol} shows the flux cancellation with higher temporal resolution with one hour cadence. Time runs from the top left to the bottom right panel showing  slow change of the magnetic field flux including subsequent disappearance of negative polarity (blue contours) and persistence of the positive polarity (red contours) followed by cooling of the atmosphere and shrinking the brightened area. As we can see in the last panel of this sequence the final disappearance is extremely rapid happening in less than half an hour. The last panel was taken half an hour apart to the second last one, which shows a strong brightening going along with the disappearance of the negative polarity. Thus Figures~\ref{cancellation} and \ref{dis_evol} clearly demonstrate that the cancellation of magnetic flux of opposite polarities within a coronal hole leads ultimately to the disappearance of brightened areas and thus to the increase of the coronal hole area.
   
\section{Interplanetary Medium and Geospace}\label{IP_sec}

 The solar wind from the coronal hole is associated with a disturbance classified as a moderate geomagnetic storm \citep{Gonzalez1994}, registered with the disturbance storm time geomagnetic index $D_{\rm st}$, with a value of $D_{\rm st} = -55$\,nT on 28~September 2017 at 07:00\,UT. Non-definitive data are available at \burl{http://wdc.kugi.kyoto-u.ac.jp/dst_realtime/201709/index.html}. The planetary $K$ geomagnetic index values were 6 and 7, from 27~September 2017 evening to 28~September 2017 in the morning. The NOAA maximum class was G3. We also note that there was a filament eruption close to AR~12681 on 25~Sep\-tem\-ber 2017 at 15:00\,UT and another larger eruption in the eastern limb, equidistant from ARs~12682 and~12683, on 25 September at 22:15\,UT.

 We used different interplanetary medium parameters obtained by the instrument suite onboard the spacecraft \textit{Advanced Composition Explorer}
 \citep[ACE,][]{Stone1998}, such as charge state abundance parameters provided by the \textit{Solar Wind Ion Composition Spectrometer} \citep[SWICS,][]{Gloeckler1998}. We also used WIND data \citep{Ogilvie_WIND1998} provided by the instrument \textit{Solar Wind Experiment} \citep[SWE,][]{Ogilvie1995}. The data comprise the components of the interplanetary magnetic field, proton density, proton bulk speed and a number of subproducts including density of He$^{2+}$.
 
 The coronal hole reached the central meridian on 24~September 2017. Therefore, solar wind from this region is expected about 2 to 5 days later. For this reason, we plot the solar wind parameters from 26~September 2017 onwards, which corresponds to Day of Year (DoY) 269. The velocity profile (top panel) shows a typical behaviour for a coronal hole speed.
 
\begin{figure}
   \centerline{\includegraphics[width=0.75\textwidth,clip=]{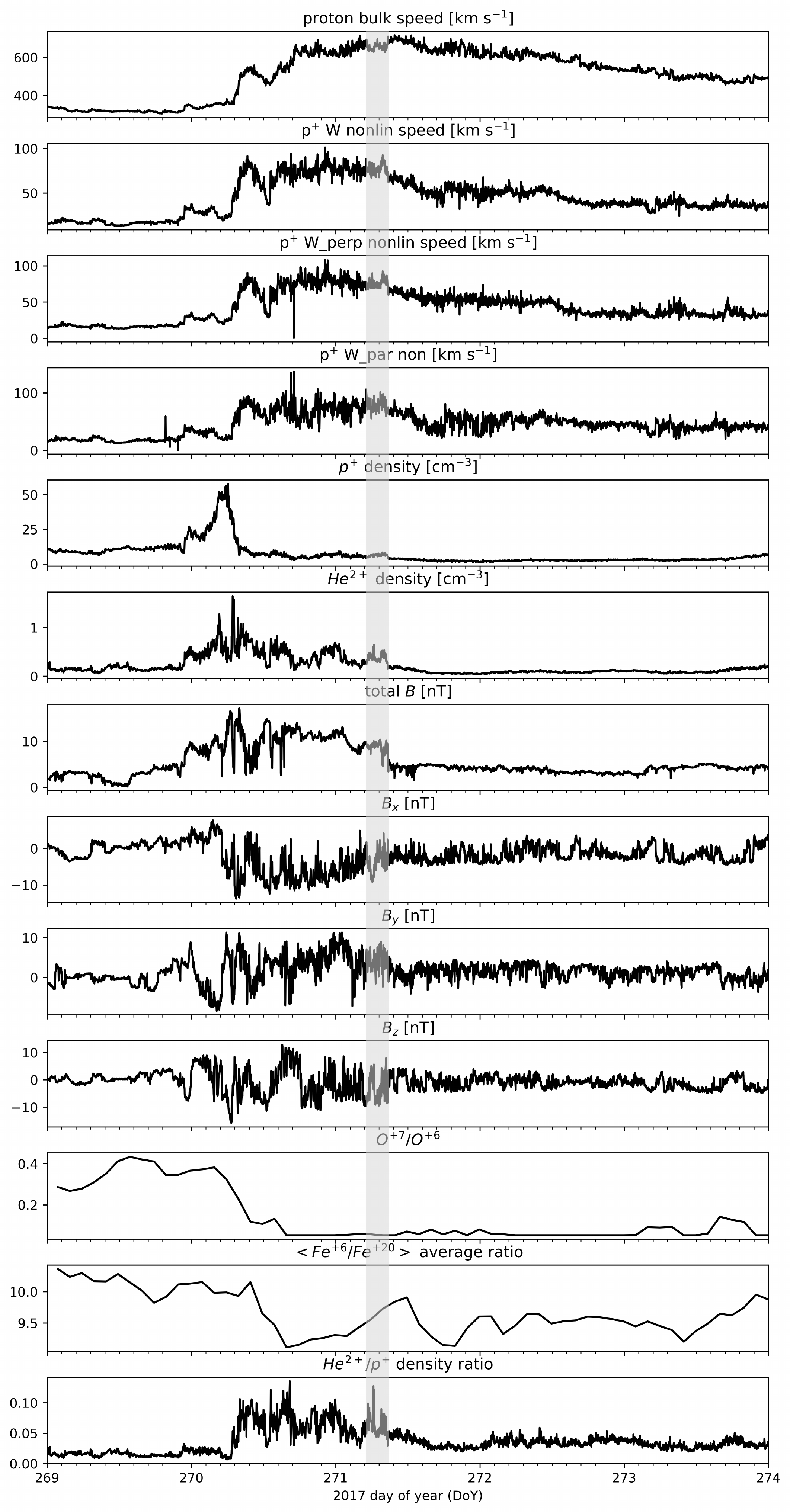}} 
              \caption{Interplanetary medium parameters. {\it From top to bottom:} proton bulk speed, proton non-linear thermal speed p$^+$W and its perpendicular p$^+$W$_{\rm perp}$ and parallel p$^+$W$_{\rm par}$ components, proton density p$^+$, He$^{2+}$ density, total interplanetary magnetic field $B$ and its three components $B_{\rm x,y,z}$, oxygen ratio O$^{7+}$/O$^{6+}$, iron ratio, and the He-to-proton density ratio He$^{2+}$/p$^+$. The shaded region shows the transient.}
  \label{IP_fig}
  \end{figure}

 In Figure~\ref{IP_fig}, a variety of interplanetary parameters are plotted in the next panels, such as proton bulk speed, total proton nonlinear thermal speed p$^+$W split into the components perpendicular p$^+$W$_{\rm perp}$ and parallel p$^+$W$_{\rm par}$ to the magnetic field orientation, proton number density p$^+$, He$^{2+}$ number density, the total and its three geocentric-solar-magnetospheric coordinate system (GSM) components of the interplanetary magnetic field, and different charge state ratios, such as O$^{7+}$/O$^{6+}$, the average Fe charge state $Q_{\rm Fe}$, and the He$^{2+}$/p$^+$ density ratio. 
 
 A peak in proton density, somewhat before the peak on bulk speed, may correspond to the corotating interacting region, as shown in Figure~\ref{IP_fig}. The interplanetary magnetic field from WIND also shows a typical structure for a high speed stream with $B_{\rm z}$ polarity corresponding to that of the coronal hole.

 Similarly to \citet{Cid2016}, we used a range of thresholds to discriminate solar wind abundance composition, which can help to distinguish non-CME structures in the solar wind, and more specifically, a high speed stream from this particular coronal hole. We show the charge states as the ratio O$^{7+}$/O$^{6+}$ and Fe average charge state $Q_{\rm Fe}$. Solar wind associated to the fast stream from the coronal hole exhibits values O$^{7+}$/O$^{6+} < 0.2$ during the time interval with large solar wind speed and fluctuating $B$ and $V$, which are in accordance of typical values. Also during this interval starting on DoY 270 about 17\,h, $Q_{\rm Fe}$ is below 10, but an increase is observed in the morning of DoY 271. The average Fe charge state can be used as a measure of the evolutionary properties in the far corona and therefore, are an excellent signature for identifying interplanetary coronal mass ejections (ICMEs). Generally, ICMEs also contain an enhanced ratio of He$^{2+}$/p$^+$, which is above 0.06 late on DoY 270 and early on DoY 271 but dropping to typical solar wind values at the rear boundary of the shadowed area in Figure~\ref{IP_fig}. Indeed, this boundary appears as a discontinuity in several solar wind plasma parameters similar to a reversal shock after a small transient with different signatures when compared to the surrounding solar wind. In the shadowed grey area, fluctuations in solar wind speed and interplanetary magnetic field (IMF), at least in $B_{\rm x}$, disappear, and a smooth variation is present. Moreover, in this region, density and temperature are not anti-correlated, as it is commonly observed in interaction regions. All these signatures suggest that material from a small transient is reaching 1 AU, probably mixed with material from the coronal hole. 

\section{Discussion}\label{disc_sec}

 Coronal hole area can suffer changes during its lifetime. In \citet{Heinemann2018a}, the decay phase of a coronal hole manifests as a decrease of the area, plus lower than average solar wind speed. A fully developed coronal hole exhibits speeds of $600 - 720$\,\kms\ \citep{Heinemann2018a}, which corresponds to the presented case here. The local decrease of the coronal hole area shown here is attributed to the described flux emergence instances.

 Flux emergence events as origin of space weather occurrences have been reported, as the case on, \textit{i.e.}, \citet{Palacios2015}. This coronal hole, with predominant positive polarity, suffers some flux emergence events, leading to area reductions. Bipolar flux emergence is significant; however, the flux is quite unbalanced between the two polarities, since the positive flux is always larger than the negative. Flux emergence rates are very similar to those in \citet{Palacios2015}. In \citet{Navarro2019}, the simulation of a blob emerging in a coronal hole leads to reconfiguration of the magnetic field and flows; however, the counterpart in solar chromospheric-coronal imaging (in EUV or X-ray imaging) is not simulated. 
 
  There are significant differences between ROI\,1 and ROI\,2 emergence events. The former corresponds to a more monolithical and fast emergence happening in the center of the coronal hole. In ROI\,2, there is a long succession of emergence and cancellation events in that area, with an emergence rate three times lower. We set an hypothesis scenario for flux emergence in the boundary of the coronal hole, where flux emerges and cancels in the boundary of the coronal hole, since its magnetic regime changes from more unipolar to areas more balanced in magnetic flux.
 
  The DAVE method for computing horizontal flow speeds shows the highest values in these EFRs, as most prominently, in the EFR at ROI\,1. Interestingly, in ROI\,2, a number of negative polarities are apparently flow sinks, which might explain that some of these negative polarity features are observed as long-lived structures. The expansion velocities tracked with DAVE are similar to other cases of medium-sized flux emergence, \textit{e.g.}, in \citet{Palacios2012}, where \textsf{Local Correlation Tracking} \citep[LCT,][]{November1988} yielded around $\sim 0.5$\,\kms\ in a 20-min dataset. Considering the difference of methods but taking into account the long-time series in the presented cases, horizontal speed values might be quite large compared to a short-time series equivalent.
 
 In addition to this, flux emergence may lead to chromospheric plasma being accelerated into a region of a vertical magnetic field of the coronal hole. The subsequent increase in gas pressure induces the ejection of plasma into the low-density corona while being heated to coronal temperatures by heating processes at the site of flux emergence.
 
 On the other hand, magnetic flux cancellation via reconnection is thought to play a major part and contribution to the creation of dynamic phenomena, \textit{e.g.}, spicules and subsequent atmospheric heating, explaining thus the brightening of the area \citep[\textit{e.g.}][]{Samanta890}. Moreover, one particular cancellation  event occurs faster than the emergence on time-scales below half an hour in ROI\,2.
 
\section{Conclusion}\label{concl_sec}

 Coronal holes and related fast solar wind are important ingredients in predicting evolution of space weather as well as constituents influencing it. In this contribution we illustrate that the evolution of the size of a coronal hole can be highly dynamic and that small-scale flux emergence events lead to a shrinkage of the coronal hole area due to the formation of bright islands within the coronal hole. Conversely, flux cancellation can lead to subsequent cooling of the atmosphere above, and locally increase the size of the coronal hole. Flux emergence and cancellation may be leading processes in the evolution of coronal holes on medium temporal scales. Thus the dynamics of the large coronal holes are given to a good part by the dynamics of the underlying small-scale photospheric magnetic fields. 
 
 This opens up interesting research avenues for the future such as the detailed as well as statistical investigation of such events and their relationship to the general evolution of coronal holes.
 
 The detected flux emergence events exhibit different behaviours in ROI\,1 and ROI\,2, which may be explained due to different localization at the center and at the boundary of the coronal hole, respectively. \textit{In-situ} measurements show features that might be compatible with the hypothesis of the photospheric flux emergence and the growth of moderate activity patches in the coronal hole, however other possibilities may be not ruled out.

\appendix  
 
The AIA 193\,\AA\ images used for the supplementary movie were processed by the noise adaptive fuzzy equalization method \citep[NAFE,][]{Druckmueller2013} to enhance visibility of the fine structures. The two free parameters of the code are $\gamma$ and \textit{w}. They are used to control the brightness and level of enhancement of the processed image. For a detailed mathematical explanation of these parameters see  \citet{Druckmueller2013}. In our case, we chose $\gamma = 2.6$ and \textit{w} = 0.25 for all processed images.

\begin{acks}[Acknowledgements]
J.P.\ acknowledges support from Leibniz-Institut f\"{u}r Sonnenphysik (KIS) on funding, computational resources and material for the creation and preparation of this manuscript. This research project received funding from the FWF under project grant P27800. P.G., S.J.G.M., and J.K.\ acknowledge project VEGA 2/0048/20. This work is part of a collaboration between AISAS and AIP supported by the German Academic Exchange Service (DAAD) under project No.\ 57449420. C.D., C.K., I.K., and M.V.\ acknowledge support by grant DE~787/5-1 of the Deutsche Forschungsgemeinschaft (DFG). M.T.\ acknowledges funding by the Austrian Space Applications Programme of the Austrian Research Promotion Agency FFG (859729, SWAMI). The support by the European Commission's Horizon 2020 Programs under grant agreements 824135 (SOLARNET) and 824064 (ESCAPE) are highly appreciated. Thanks to Y. Hanaoka for providing data from the Infrared Stokes-Polarization Full-Disk Images from the Solar Flare Telescope. Moreover, the authors want to acknowledge SDO/AIA and SDO/HMI Data Science Centers and Teams. Data were obtained during a joint GREGOR campaign with support by Hinode, IRIS, VTT, plus Chrotel. The 1.5-meter GREGOR solar telescope was built by a German consortium under the leadership of KIS with AIP, and MPS as partners, and with contributions by the IAC and ASU. Hinode is a Japanese mission developed and launched by ISAS/JAXA, with NAOJ as domestic partner and NASA and STFC (UK) as international partners, which is operated by these agencies in co-operation with ESA and NSC (Norway). We acknowledge data use from ACE and WIND spacecraft instruments, and to STEREO as well. We acknowledge data use from WDC from Geomagnetism, Kyoto, and LMSAL SolarSoft. NASA Astrophysics Data System (ADS) has been used as bibliographic engine.\par
\end{acks}

\bibliographystyle{spr-mp-sola}
\bibliography{./jp_ch.bib}

\end{article} 

\end{document}